\documentclass[prl,twocolumn,amsmath,amssymb,floatfix,showpacs,superscriptaddress]{revtex4-2}
\usepackage{graphicx}% Include figure files
\usepackage{amsmath}% Include figure files
\usepackage{placeins}
\usepackage{bm}
\usepackage[dvipsnames]{xcolor}
\usepackage[caption=false]{subfig}
\usepackage{hyperref}

\begin{document}

\title{Shape-induced pairing of spheroidal squirmers}
\author{Ruben Poehnl}
\affiliation{Department of Mechanical Engineering, University of Hawai’i at 
M{\=a}noa, 2540 Dole Street, Holmes Hall 302, Honolulu, Hawaii 96822, USA}

\author{William E. Uspal}
\email{uspal@hawaii.edu}
\affiliation{Department of Mechanical Engineering, University of Hawai’i at 
M{\=a}noa, 2540 Dole Street, Holmes Hall 302, Honolulu, Hawaii 96822, USA}

\begin{abstract}
The “squirmer model” is a classical hydrodynamic model for the motion of interfacially-driven microswimmers, such as self-phoretic colloids or volvocine green algae. To date, most studies using the squirmer model have considered spherical particles with axisymmetric distribution of surface slip. Here, we develop a general approach to the pairing and scattering dynamics of two spheroidal squirmers. We assume that the direction of motion of the squirmers is restricted to a plane, which is approximately realized in many experimental systems. In the framework of an analytically tractable kinetic model, we predict that, for identical squirmers, an immotile “head-to-head” configuration is stable only when the particles have oblate shape and a non-axisymmetric distribution of surface slip. We also obtain conditions for stability of a motile “head-to-tail” configuration: for instance, the two particles must have unequal self-propulsion velocities. Our analytical predictions are compared against detailed numerical calculations obtained using the boundary element method.
\end{abstract}

\date{\today}
\maketitle

\section{Introduction}
Self-assembly \cite{sanchez2012,aubret2018,boymelgreen2018,Arora21}, clustering \cite{palacci2013,Buttinoni13,pohl14,cates2015,hokmabad2022}, and particle motility alignment \cite{bricard13, yan2016, kaiser2017flocking, han2020, Zhang21} are among the most discussed topics in the active matter community. In each of these phenomena, collective behavior emerges from non-equilibrium interactions between \textcolor{black}{self-motile} microscopic particles. \textcolor{black}{These particles usually self-propel through liquid, making hydrodynamic interactions -- interactions mediated by flow in the suspending medium -- an important and ubiquitous non-equilibrium effect \cite{martinez2018advances}.}    

% Not only do these phenomena provide test cases for understanding how microscopic properties are connected with emergent collective behavior, but they are also essential for many envisioned real world applications. 

\textcolor{black}{A broad class of synthetic and biological microswimmers propel themselves by driving flow within a thin layer at the fluid/solid interface. For instance, ciliated microorganisms are covered by a thin carpet of thread-like appendages that beat in  a coordinated fashion.} The  
squirmer model, first introduced by Lighthill \cite{Lighthill52} and Blake \cite{Blake71}, was originally  developed to describe the motion of ciliated, spherical micro-organisms \cite{pedley2016spherical}. \textcolor{black}{In the simplest version of this model, the  detailed, time-dependent motion of the cilia is coarse-grained as a prescribed steady tangential slip velocity.  This slip velocity  provides the interfacial actuation (\textit{i.e.}, thrust) needed for self-propulsion. Additionally, the slip velocity drives flow in the surrounding solution, leading to long-ranged hydrodynamic interactions between the squirmer and other objects in the solution. These features have made the squirmer model a popular approach for understanding the flow-mediated interactions between swimming microorganisms, as well as between microorganisms and bounding surfaces \cite{ishikawa2006,ishikawa2007diffusion,drescher09,llopis2010hydrodynamic, ishimoto2013squirmer,li2014hydrodynamic,darveniza22}. For instance, Ishikawa \textit{et al.} exhaustively cataloged the collision and scattering dynamics of squirmer pairs \cite{ishikawa2006}.  As another example, various studies have addressed nutrient uptake (i.e., feeding) of microorganisms in the framework of the squirmer model \cite{magar03,michelin2011}.  }

 \textcolor{black}{Since its original development, the squirmer model has } found applications well beyond its initial purpose. For instance, \textcolor{black}{synthetic} active colloids \textcolor{black}{driven by self-generated gradients of a thermodynamic variable (e.g., temperature, chemical potential, or electrical potential \cite{wang2006bipolar, howse2007self,jiang2010active,bregulla2019flow})} can often be approximated as ``effective squirmers.'' Instead of resolving the propulsion mechanism in detail, a slip velocity on the surface of the particle is prescribed \cite{Popescu18}. The slip determines two major swimming properties, speed $U_s$ and stresslet $\mathbf{S}$ \cite{Lauga16, Stone96}. As an example that justifies this approach, it was recently observed that metallo-dielectric Janus discoids, energized by AC fields, tend to form ``head-to-head'' bound pairs \cite{katuri22}. Modeling of the propulsion mechanism (induced charge electrophoresis \cite{squires06,kilic11}) revealed that hydrodynamic interactions dominated interactions between particles, i.e., the particles behaved as effective squirmers.

One microscopic property that has proven to be  important \textcolor{black}{in active matter} is particle geometry \cite{brooks2018shape,sharan2021fundamental, diwakar2022,ganguly2023going}. Shape can impact the swimming speed of an active particle, the rate of working, and the flow field sourced by the particle \cite{theers16, michelin2017geometric, yariv2019self, daddi2021optimal,poehnl21,zantop2022emergent}. Collisions between rod-like particles can lead to nematic ordering in an active suspension \cite{bar2020self}. \textcolor{black}{In view of the importance of shape, various studies have considered non-spherical squirmers -- usually prolate spheroids \cite{ishimoto2013squirmer,theers16,felderhof2016stokesian,leshansky2007frictionless}. For instance, Ishikawa and Hoto modeled the paramecium \textit{P. caudatum} with a prolate spheroid acutuated by interfacial slip. The slip was assumed to be a superposition of five harmonic functions of the elevation angle \cite{ishikawa2006interaction}.}    \textcolor{black}{In an effort to fully generalize the squirmer model to both prolate and oblate spheroids, we recently developed and characterized a complete set of orthogonal, axisymmetric squirming modes in spheroidal coordinates \cite{poehnl20}. We found that the  odd-numbered squirming modes contribute to the self-propulsion velocity, while the even-numbered contribute to the the stresslet.} 

For interfacially-driven microswimmers, a second means of controlling their motion is offered by breaking symmetries of the slip velocity. This symmetry breaking can imposed, as when a self-phoretic particle is fabricated with non-axisymmetric surface chemistry \cite{archer2015,Lisicki18}, or can emerge \textit{in situ}, due to effects of confinement \cite{kilic11,uspal15} or symmetry-breaking fields \cite{squires06,popescu2018chemotaxis}. So far, applications of the squirmer model have mostly been restricted to axisymmetric slip, although more general slip has been considered for spherical squirmers \cite{Ghose14,Pak14,felderhof2016,pedley16}. \textcolor{black}{For instance, Burada \textit{et al.} consider the far-field interaction of two spherical squirmers with a chiral distribution of slip. They find that these spheres can exhibit oscillatory ``bounded states'' in which they orbit around a common average trajectory \cite{burada2022}.}

\section{Theory}
In this work, we develop a framework to study the consequences of particle shape and non-axisymmetry of the surface slip for interactions between interfacially-driven microswimmers. \textcolor{black}{We develop analytical predictions in a far-field, ``point-particle'' model, building on the Saintillan-Shelley kinetic theory of microswimmers \cite{saintillan2007orientational,saintillan2008,Saintillan13,lushi2013modeling}. Our analytical predictions are supported by high resolution numerical calculations using the squirmer model, which resolve the finite size of the particle and near-field hydrodynamic interactions.} We show that both non-spherical shape and breaking of the axisymmetry are necessary conditions to form stable ``head-to-head'' bound pairs. These immotile bound states are held together by (far-field) hydrodynamic interactions. Similarly, we find that motile ``head-to-tail'' bound pairs can be stable only when the particles are non-spherical (although they can be axisymmetric.) Overall, we find good agreement between theory and numerics, suggesting that our framework offers a promising approach for studying self-organization in hetereogeneous active suspensions.

\subsection{Minimal model} 
We model swimmer $\alpha \in \{1,2\}$ as a point-like particle with swimming direction $\hat{d}^{(\alpha)}$ and self-propulsion velocity $U^{(\alpha)}_s \geq 0$. Swimmers are coupled by the flows they generate. The velocity of swimmer $\alpha$ is 
\begin{equation}
\mathbf{U}^{(\alpha)} = U_{s}^{(\alpha)} \, \hat{d}^{(\alpha)} + \mathbf{u}(\mathbf{x}_{\alpha}).
\end{equation}
In the second term, the swimmer is advected by the ambient flow, evaluated at its position $\mathbf{x}_{\alpha}$. (The finite size of the swimmer is neglected.) For the rotation of the swimmer, we write the Jeffery equation \cite{Saintillan13}
\begin{equation}
\label{eq:jeffery}
\dot{\hat{d}}^{(\alpha)} = \left(\mathcal{I} - \hat{d}^{(\alpha)} \hat{d}^{(\alpha)} \right) \cdot \left( \Gamma_{\alpha} \mathbf{E}(\mathbf{x}_\alpha) + \mathbf{W}(\mathbf{x}_\alpha)  \right) \cdot \hat{d}^{(\alpha)}.
\end{equation}
Here, $\Gamma_{\alpha}$ is a shape parameter that is zero for a sphere, positive for a prolate spheroid that has its major axis aligned with $\hat{d}^{(\alpha)}$, and negative for an oblate spheroid that has its minor axis aligned with $\hat{d}^{(\alpha)}$. The tensors $\mathbf{E}(\mathbf{x_\alpha})$ and $\mathbf{W}(\mathbf{x_\alpha})$ are the rate-of-strain and vorticity, respectively, evaluated at $\mathbf{x}_{\alpha}$, where $\mathbf{E} = \frac{1}{2} \left(\nabla \mathbf{u} + \nabla \mathbf{u}^T\right)$ and $\mathbf{W} = \frac{1}{2} \left(\nabla \mathbf{u} - \nabla \mathbf{u}^T\right)$.
${\mathcal{I}}$ is the identity tensor.

To model swimmer-generated flow, we associate an active ``stresslet'' with each swimmer. In general, the stresslet provides the slowest decaying contribution of a force-free, \textcolor{black}{rigid} microswimmer to the surrounding flow field. It can be obtained from the surface traction \cite{ishikawa2006, kim2013microhydrodynamics}:
\begin{equation}
\label{eq:stresslet_integral}
S_{ij}^{(\alpha)} = \frac{1}{2} \int_{\Sigma_\alpha} [\sigma_{ik} \textcolor{black}{\hat{n}}_k x_j + \sigma_{jk} \hat{n}_k x_i] \, dS - \frac{1}{3} \int_{\Sigma_\alpha}  \sigma_{lk} \hat{n}_k x_l \, dS \, \delta_{ij}.
\end{equation}
The integral is taken over the surface $\Sigma_\alpha$ of particle $\alpha$, $\hat{n}$ points from the surface of the particle into the surrounding fluid, and $\bm{\sigma} = - p {\mathcal{I}} + \mu (\nabla \mathbf{u} + \nabla \mathbf{u}^T )$ is the stress tensor for a Newtonian fluid. Here, $p(\mathbf{x})$ is the pressure and $\mu$ is the dynamic viscosity of the fluid. The velocity field due to a stresslet located at the origin is given by:
\begin{equation}
u_i = \frac{1}{8 \pi \mu } \left( \frac{x_i \delta_{jk}}{r^3} - \frac{3 x_i x_j x_k}{r^5} \right) S_{jk}^{(\alpha)},
\end{equation}
where $r$ is distance from the origin and $x_i$ is a location in the fluid. For a swimmer with an \textit{axisymmetric} surface actuation, the stresslet can be written as \cite{Saintillan17}
\begin{equation}
\label{eq:axisym_stresslet}
\mathbf{S}^{(\alpha)} = \sigma_0^{(\alpha)} \left( \hat{d}^{(\alpha)} \hat{d}^{(\alpha)} - \frac{\mathcal{I}}{3} \right).
\end{equation}
The sign of $\sigma_0^{(\alpha)}$ determines the ``pusher'' ($\sigma_0^{(\alpha)} < 0$) or ``puller'' ($\sigma_0^{(\alpha)} > 0$) character of the swimmer.

\begin{figure}[t!]
\centering
\includegraphics[width=8.6cm]{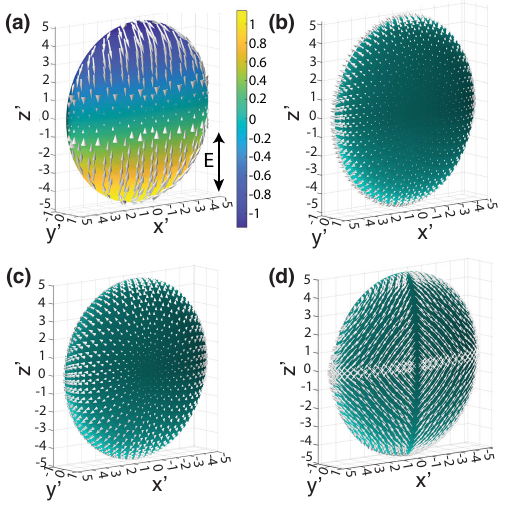}
\caption{\label{Fig:slips} (a) An oblate particle that self-propels in the presence of an AC electric field (black arrow) by ICEP. The background color indicates the electrostatic potential (arbitrary units) on the particle surface for the half-period in which the field is pointing in the negative z' direction. White arrows show the surface slip, which is non-axisymmetric. (b) Distribution of slip for the first squirming mode $B_1$. (c) Distribution of slip for $B_2$. (d) Non-axisymmetric slip for $\tilde{B}$, following the definition in Eq. \ref{eq:nasym_slip}.
In all panels, \textcolor{black}{$r_e = 5$}.}
\end{figure}

However, not all microswimmers have axisymmetric actuation. For instance, consider metallo-dielectric particles that are energized by an AC electric field and swim via induced charge electrophoresis (ICEP) \cite{squires06}. The applied field can break axisymmetry. Thus, we consider a more general stresslet, written in a frame aligned with the principal axes $\hat{c}$, $\hat{d}$, and $\hat{e}$ of $\mathbf{S}^{(\alpha)}$:
\begin{equation}
\label{eq:ICEPstresslet}
\mathbf{S}^{(\alpha)} = S_{cc}^{(\alpha)} \hat{c}\hat{c}  + S_{dd}^{(\alpha)} \hat{d}\hat{d} + S_{ee}^{(\alpha)} \hat{e} \hat{e}, \end{equation}
with $\textrm{tr}(\mathbf{S}^{(\alpha)}) = 0$. Since $\mathbf{S}^{(\alpha)}$ is symmetric \textcolor{black}{and real-valued}, its principal axes are orthogonal, and we define $\hat{c} \times \hat{d} = \hat{e}$. \textcolor{black}{This form of the stresslet tensor is generic. However, for} simplicity, we \textcolor{black}{make the assumption} that the direction of propulsion of an isolated particle is $\hat{d}$, \textit{i.e.}, aligned with a principal axis. This assumption is realized by an ICEP particle with spheroidal shape and axisymmetric metal coverage, swimming in unbounded solution (Fig. \ref{Fig:slips}(a)). If the electric field is in the $\hat{z}$ direction and the particle axis of symmetry is given by $\hat{d}$, the particle will rotate such that $\hat{d}$ is perpendicular to $\hat{z}$ \cite{squires06,diwakar2022}. After rotation, the particle will swim strictly in $\hat{d}$ with a stresslet tensor in the form of Eq. \ref{eq:ICEPstresslet}. A  detailed technical discussion of \textbf{S} and the assumption concerning $\hat{d}$ is provided in the Supplemental Material (SM) \footnote{See Supplemental Material at \url{http://link.aps.org/supplemental/XXX} for technical discussion of the stresslet tensor, detailed mathematical derivations of results given in the main text, details concerning implementation of the spheroidal squirmer model, and numerical results obtained with the spheroidal squirmer model. The Supplemental Material also contains Refs. \citenum{kilic11,poehnl20,katuri22,dassios1994generalized}.}. Additionally, we note that Eq. \ref{eq:ICEPstresslet} reduces to Eq. \ref{eq:axisym_stresslet} when $S_{dd}^{(\alpha)} = 2/3 \sigma_0$ and $S_{cc}^{(\alpha)} = S_{ee}^{(\alpha)} = -\sigma_0/3$. 

In the following, we restrict our consideration to two swimmers moving in the $xy$ plane, and study conditions for obtaining stable bound states. The instantaneous configuration of the system is specified by the center-to-center distance $d$ and the angles $\phi_1$ and $\phi_2$, where $\phi_\alpha$ is the angle between $\hat{d}^{(\alpha)}$ and a fixed axis, chosen as the x-axis (see Fig. 1 in the \textcolor{black}{SM}). \textcolor{black}{We assume that $\hat{c}^{(\alpha)}$ and  $\hat{d}^{(\alpha)}$ lie within the xy plane.} For convenience, we specify that swimmer 1 is instantaneously at $\mathbf{x}_{1} = (0,0,0)$. Swimmer 2 has position $\mathbf{x}_{2} = (x,y,0)$. We construct $\dot{x}$, $\dot{y}$, $\dot{\phi}_1$, and $\dot{\phi}_2$ as functions of $x$, $y$, $\phi_1$, and $\phi_2$, and look for fixed point configurations at which these functions evaluate to zero, representing a bound state. For simplicity, we consider only bound states in which the propulsion axes are aligned with the center-to-center axis. 

The point-particle model is analytically tractable and yields a wealth of predictions. However, we wish to compare these predictions against numerical results that account for finite size and do not truncate the particle-generated flow field to the leading-order term. 

%To evaluate the stability of a bound pair, we perform linear stability analysis. %we compute the Jacobian $\frac{\partial \dot{\mathbf{f}}}{\partial \mathbf{f}}$

\subsection{Squirmer model} 
We consider $N \in \{1,2\}$ spheroidal particles in unbounded Newtonian fluid. Following Ref. \citenum{poehnl20}, for particle $\alpha$, we take the length of the semi-axis of symmetry to define $b_y^{(\alpha)}$, and the length of the other semi-axes to define $b_x^{(\alpha)}$. Thus, each particle has an aspect ratio $r_e^{(\alpha)}$ defined by {$r_e \equiv b_x/b_y$, with $r_e < 1$ for an prolate spheroid, $r_e = 1$ for a sphere, and $r_e > 1$ for an oblate spheroid.} The quantity $\Gamma_\alpha$ is related to $r_e^{(\alpha)}$ by $\Gamma = (1 - r_e^2)/(1 + r_e^2)$. %The characteristic size of the particle, $L_0^{(\alpha)}$, is chosen as $\textrm{max}(b_x,b_z)$.
The characteristic size of the particle, $L_0^{(\alpha)}$, is chosen as $b_y^{(\alpha)}$.

The center of particle $\alpha$ is located at $\mathbf{x}_{\alpha}$. The fluid velocity $\mathbf{u}(\mathbf{x})$ is governed by the Stokes equation
$-\nabla p + \mu \nabla^2 \mathbf{u} = 0$ and the incompressibility condition $\nabla \cdot \mathbf{u} = 0$. On the surface $\Sigma_{\alpha}$ of particle $\alpha$, the velocity obeys $\mathbf{u} = \mathbf{U}^{(\alpha)} + \bm{\Omega}^{(\alpha)} \times (\mathbf{x} - \mathbf{x}_{\alpha}) + \mathbf{v}_{s}^{(\alpha)}(\mathbf{x})$. Additionally, $|\mathbf{u}| \rightarrow 0$ as $|\mathbf{x}| \rightarrow \infty$. Each particle is force-free and torque-free: $\int_{\Sigma_{\alpha}} \bm{\sigma} \cdot \bm{\hat{n}} \, dS = 0$ and $\int_{\Sigma_{\alpha}} (\mathbf{x} - \mathbf{x}_{\alpha}) \times \bm{\sigma} \cdot \bm{\hat{n}} \, dS = 0$. 

For each swimmer, the slip $\mathbf{v}_s^{(\alpha)}$ is fixed in a frame attached to the swimmer. It is specified via a set of squirming mode amplitudes. In previous work, we generalized the axisymmetric squirming modes to spheroidal particles. The amplitudes are denoted by $B_i^{(\alpha)}$, with $i \geq 1$ \cite{poehnl20}, and here are assumed to be given in units of an arbitrary characteristic velocity. The first two modes are shown in Fig. \ref{Fig:slips}(b) and (c). Here, in order to break axisymmetry, we develop a new squirming mode $\tilde B^{(\alpha)}$ inspired by the slip profile of ICEP particles. This squirming mode has slip distribution
\begin{align}
\label{eq:nasym_slip}
 \mathbf{v}_{s}^{(\alpha)}(\mathbf{x})= \tilde B^{(\alpha)}  [&\text{sign}(\textcolor{black}{z'})(\cos(\varphi)\hat{e}_\xi-\hat{e}_\varphi(\hat{e}_\xi\cdot\hat{e}_{\textcolor{black}{z'}}))\\\nonumber
 &-\text{sign}(\textcolor{black}{x'})(\sin(\varphi)\hat{e}_\xi-\hat{e}_\varphi(\hat{e}_\xi\cdot\hat{e}_{\textcolor{black}{x'}})) ]\cdot H(\textcolor{black}{y'}),
\end{align}
shown in Fig. \ref{Fig:slips}(d). Here, $H(\textcolor{black}{y'})$ is the step function, and $\hat{e}_\varphi$ and $\hat{e}_\xi$ are two surface tangential basis vectors in a particle-centered spheroidal coordinate system. \textcolor{black}{(The prime symbol is used to distinguish the coordinate system in Fig. \ref{Fig:slips} from the coordinate system used for studying pair interactions.)} %Only $\hat{e}_\xi$ is axisymmetric, and thus the above defined slip mode is tangential but non-axisymmetric.

We briefly discuss the properties of a single squirmer. From solution of the governing equations, we obtain $U_s$ and $\mathbf{S}$ for a given $r_e$ and set of squirming mode amplitudes. Due to the linearity of the Stokes equation, the contribution of each squirming mode can be calculated individually and superposed. For the axisymmetric modes, Fig. 2 in the \textcolor{black}{SM} shows how the $B_i$ contribute to $U_s^{(\alpha)}$ and $\sigma_0^{(\alpha)}$. For the non-axisymmetric mode, we show $S_{cc}^{(\alpha)}$, $S_{dd}^{(\alpha)}$ and $S_{ee}^{(\alpha)}$ as a function of $r_e$ in \textcolor{black}{SM} Fig. 5. This squirming mode makes no contribution to $S_{dd}^{(\alpha)}$ or $U_s^{(\alpha)}$, and contributes anti-symmetrically to $S_{cc}^{(\alpha)}$ and $S_{ee}^{(\alpha)}$. 

For $N = 2$, we solve for the particle velocities numerically, using the boundary element method (BEM) \cite{pozrikidis02}. We obtain trajectories using \textcolor{black}{a rigid body dynamics engine} \cite{theers16}. For simplicity, we assume that $L_0^{(1)} = L_0^{(2)}$. (The point-particle model has no inherent length scale. Since $S_{ij} \sim L_0^3$ and $U_s \sim L_0^2$ for a squirmer, differences in size can be straightforwardly accommodated \textcolor{black}{in our model}.)

\section{Results}
\subsection{Head-to-head pairing}
 We look for fixed point solutions of the point-particle model with $(x,y,\phi_1,\phi_2) = (d_0,0,0,\pi)$.  \textcolor{black}{Through a detailed derivation in the SM,} we obtain 
\begin{equation}
\label{eq:separation_hh}
d_0 = \sqrt{\frac{-3 \left(S_{dd}^{(1)} + S_{dd}^{(2)}\right) }{8 \pi \mu \left(U_s^{(1)} + U_s^{(2)}\right)}}.
\end{equation}
Given that $U_s^{(\alpha)} > 0$, to obtain a finite separation $d >0$, it is required that $(S_{dd}^{(1)} + S_{dd}^{(2)}) < 0$. \textcolor{black}{In other words, the pair must have a net ``pusher'' character.}
In the \textcolor{black}{SM}, we present a general linear stability analysis. Here, we discuss identical swimmers, i.e., $U_{s}^{(1)} = U_{s}^{(2)}$, $\mathbf{S}^{(1)} = \mathbf{S}^{(2)}$, and $\Gamma_1 = \Gamma_2$. As conditions for stability, we obtain $\Gamma < -1/3$ and $[S_{cc}(-1 + \Gamma) + S_{dd}(1 + 2 \Gamma)] [S_{cc} (-1 + \Gamma) - S_{dd}(1 + 4 \Gamma)] < 0$, given that $S_{dd} < 0$. Notably, the requirement $\Gamma < -1/3$ corresponds to an oblate shape, recalling the discoidal particles in Ref. \citenum{katuri22}. Intriguingly, head-to-head pairing cannot be obtained for \textit{axisymmetric} swimmers (Eq. \ref{eq:axisym_stresslet}). For $S_{dd} = 2\sigma_0/3$ and $S_{cc} = S_{ee} = -\sigma_0/3$, with $\sigma_0 < 0$, the second condition reduces to $\Gamma > \textcolor{black}{-}1/9$. This cannot be reconciled with $\Gamma < -1/3$. Thus, this work completes the analysis of Ref. \citenum{katuri22}, which assumed an axisymmetric stresslet. Here, we have shown that non-axisymmetry is a necessary ingredient in the pairing observed in Ref. \citenum{katuri22}. 

To further investigate deviation from axisymmetry, we consider stresslets of the form
\begin{equation}
\mathbf{S} = \mathbf{S}_{ax} + \sigma_0 \, \delta \left(  \hat{c} \hat{c} -  \hat{e} \hat{e}\right), 
\label{eq:delta}
\end{equation}
where $\mathbf{S}_{ax}$ is equal to the right hand side of Eq. \ref{eq:axisym_stresslet}, and $\delta$ is dimensionless. We obtain $\Gamma < -1/3$ and $(-1 + 3 \Gamma (-3 + \delta) - 3 \delta) (1  + \Gamma + \delta (-1 + \Gamma) )  < 0$. Notably, these requirements are independent of $\sigma_0$ and $U_s$.  
\begin{figure}
    \centering
  \includegraphics[]{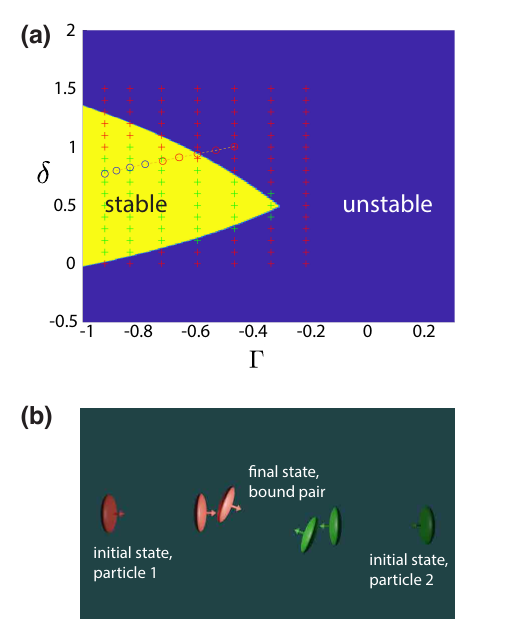}
    \caption{(a) Phase map for head-to-head pairs. The background colors show the stability predicted by \textcolor{black}{the analytical} model. Crosses represent \textcolor{black}{the results of numerical calculations} for non-axisymmetric squirmers (Eq. \ref{eq:nasym_slip}) with modes $B_1^{(\alpha)}=0.1$, $B_2^{(\alpha)}=-1$ and varying $\tilde{B}$ and $\Gamma$. Circles indicate \textcolor{black}{numerical} data for the ICEP effective squirmer model, and are connected by a line to guide the eye. \textcolor{black}{Green and blue symbols indicate pairs with with a stable bound state; red symbols represent unstable pairs.} (b) Trajectory obtained for $\Gamma = -0.835$ and $\tilde{B} = 1.68$. }
    \label{Fig:phase_head_head}
\end{figure}
In Fig. \ref{Fig:phase_head_head}(a), the background color shows the predicted phase map. We also show two types of numerical data. Crosses represent squirmers with a non-axisymmetric squirming mode. This mode introduces the perturbation $\delta$ in a controllable manner (see Fig. 5 in the \textcolor{black}{SM}). Circles show the results for an effective squirmer model for ICEP particles. Red symbols indicate pairs without a stable bound state. The theoretical and numerical results largely agree with each other. The one area of significant mismatch is for $\Gamma \approx -1$, i.e., oblate spheroids with large $r_e$. Recalling that $b_y$ was chosen as a characteristic length, oblate particles with large $r_e$ also have large $b_x$. When $b_x \gg d_0$, the point particle assumption is expected to be erroneous. In Fig. \ref{Fig:phase_head_head}(b), we show pair formation for $\Gamma = -0.835$ and $\tilde{B} = 1.68$.

\begin{figure}
    \centering  \includegraphics[width=\columnwidth]{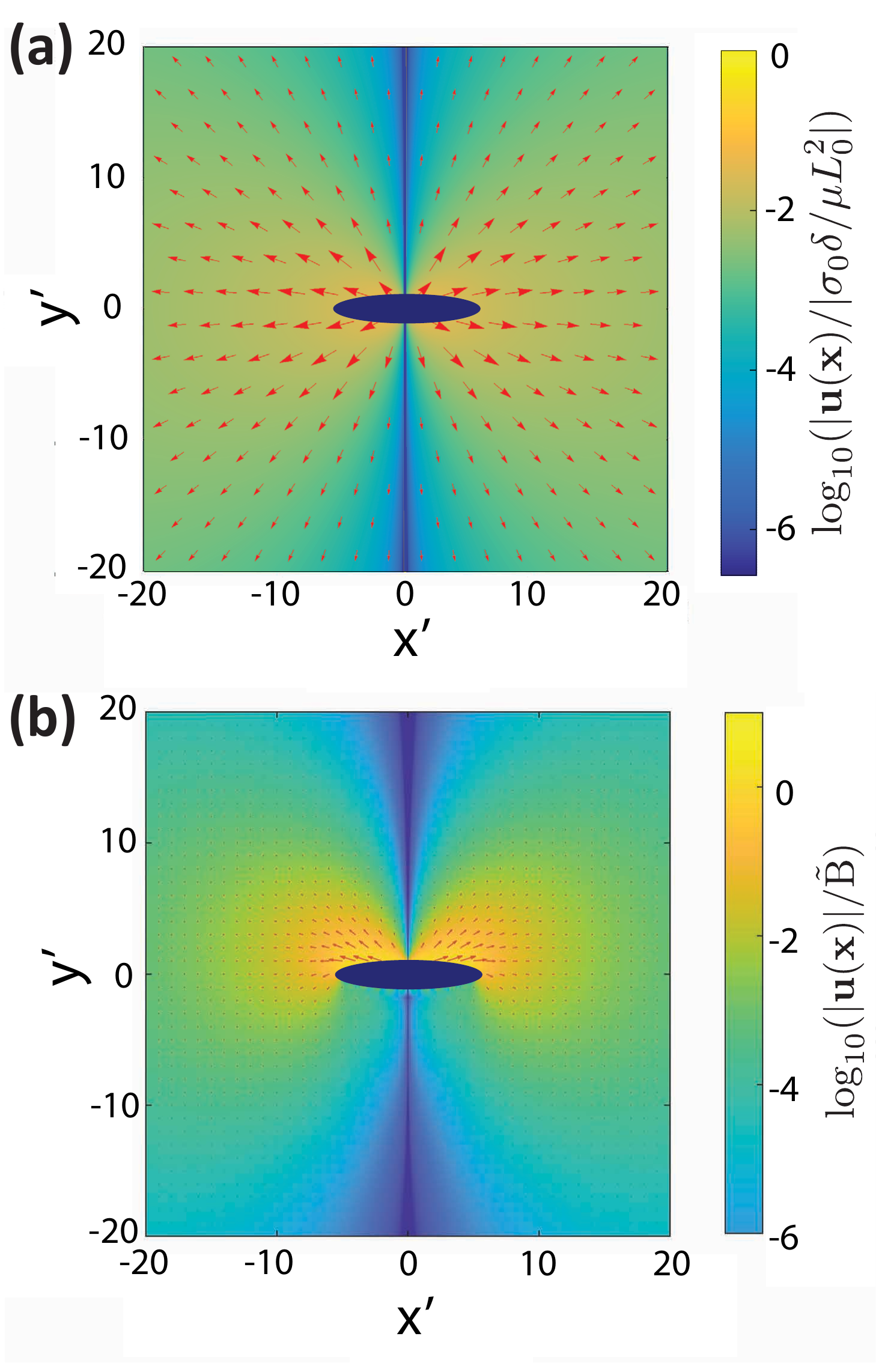}
    \caption{\textcolor{black}{(a) Flow field due to a non-axisymmetric stresslet located at the origin,  with $\hat{c}$ in the $x'$ direction and $\hat{e}$ in the $z'$ direction. The direction of the flow is shown for the case $\sigma_0 \delta < 0$. The flow is radially outward, but the magnitude is anisotropic. An oblate spheroid is shown for comparison with (b).  (b) Flow field due to the non-axisymmetric $\tilde{B}$ mode for the oblate spheroidal particle shown in Fig. \ref{Fig:slips}(d). In both (a) and (b), the flow velocity is zero at $x' = 0$.}}
    \label{fig:flow_fig}
\end{figure}

\textcolor{black}{The condition $\Gamma < -1/3$ has a straightforward physical interpretation. At the location of a particle, the rate-of-strain tensor $\mathbf{E}$ has two principal axes. Spheroidal particles tend to align their long axes with the local axis of extension \cite{jeffery1922motion,graham2018microhydrodynamics}. For an axisymmetric stresslet (Eq. \ref{eq:axisym_stresslet})  located at the origin and oriented in the x-direction, we evaluate $\mathbf{E}_{ax}$ at the position $x = d$, $y = 0$. From the eigenvalues and eigenvectors of this quantity, we find that the axis of extension is indeed in the y-direction.  Thus, the straining component of flow will tend to stabilize the orientation of oblate spheroids in a head-on collision. Furthermore, we note that $\delta$ does not appear in the condition $\Gamma < -1/3$. As a consistency check, we form the rate-of-strain tensor $\mathbf{E}_{\delta}$ for the non-axisymmetric contribution to flow (second term in Eq. \ref{eq:delta}), assuming that $\hat{d} = \hat{x}$ and $\hat{c} = -\hat{y}$. We find that it indeed evaluates to zero at $x = d$, $y = 0$. Additionally, in Fig. \ref{fig:flow_fig}(a), we plot the flow from the idealized non-axisymmetric stresslet. On the $\hat{d}$ axis (in the figure, the $y'$ axis), it evaluates to zero, which explains why $d_0$ is determined by the axisymmetric component of the stresslet (Eq. \ref{eq:separation_hh}).}

\textcolor{black}{The second requirement for linear stability,  $(-1 + 3 \Gamma (-3 + \delta) - 3 \delta) (1  + \Gamma + \delta (-1 + \Gamma) )  < 0$, is more difficult to interpret. The quantities $\Gamma$ and $\delta$ are implicated, both individually and as a product with each other. Additionally, by introducing dummy variables into the Jacobian, we have confirmed that both vorticity and transverse advection (\textit{i.e.}, motion in $y$, transverse to the center-to-center vector) contribute to this condition. Some insight can be obtained from the form of the Jacobian in SM Eq. 40. The stresslet component $S_{cc}$ appears only in off-diagonal terms that couple transversal displacements and particle rotations.  This suggests that the non-axisymmetric stresslet is important in the intricate dance in which particles simultaneously rotate to face each other and slide laterally into register, as shown in Fig. \ref{Fig:phase_head_head} (b). 
In contrast, it is known that spherical squirmers in a head-on collision are unstable to maneuvering past each other in a process involving rotations and transversal motion \cite{katuri22}. Looking at the flow for the non-axisymmetric stresslet in Fig.  \ref{fig:flow_fig}(a), some stabilizing roles of this radially outward flow may be in hindering the particles from moving past each other }\textcolor{black}{and in contributing to alignment. Regarding alignment, we recall that the magnitude and sign of the contribution of the rate-of-strain tensor to rotation is controlled by the shape parameter $\Gamma$ (Eq. \ref{eq:jeffery}). We also note that while the flow fields close to the particle can differ significantly between the idealized non-axisymmetric stresslet in Fig. \ref{fig:flow_fig}(a) and the non-axisymmetric squirming mode in Fig. \ref{fig:flow_fig}(b), far from the particle, both flow fields are radially outward.}

\begin{figure}
\centering
\includegraphics[]{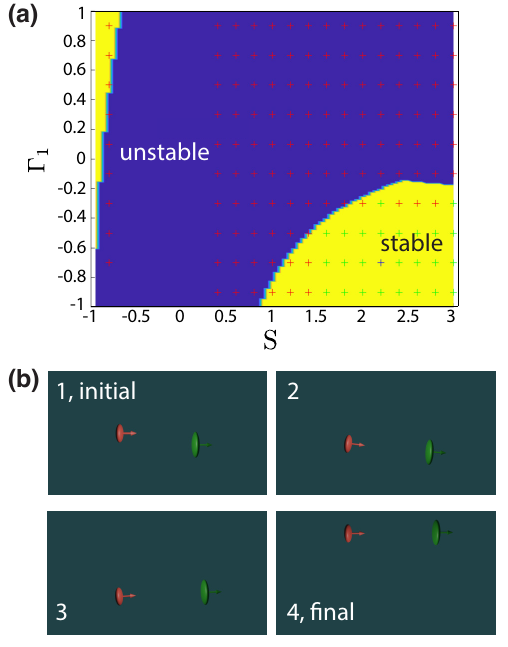}
\caption{(a) Phase map for head-to-tail pairs with $\Gamma_2=-0.8$ and $V=0.8$ and varying $S$ and $\Gamma_1$. The background colors show the stability predicted by our model and symbols represent \textcolor{black}{the results of} numerical calculations. \textcolor{black}{Green and blue symbols indicate pairs with a stable bound state; red symbols indicate pairs without a stable bound state.} (b) Snapshots of an example trajectory for $S=2.2$ and $\Gamma_1=-0.7$. This pair is represented by a blue cross in (a). The particles are initially separated by $x_\text{2,initial}=3$ and $y_\text{2,initial}=20$.}
\label{Fig:phase_head_tail}
\end{figure}

\subsection{Head-to-tail pairs}
Now we look for fixed point solutions with $(x,y,\phi_1,\phi_2) = (d_0,0,0,0)$. We obtain
\begin{equation}
 d_0 = \sqrt{\frac{3 \left(S_{dd}^{(1)} + S_{dd}^{(2)} \right)}{8 \pi \mu (U_s^{(2)} - U_s^{(1)})}}.
 \label{eq:headtotaildist}
\end{equation}
Notably, the two particles must have unequal speeds $U_s$ for $d_0 > 0$. The bound pair moves with a steady speed given by Eq. \textcolor{black}{27} in the \textcolor{black}{SM}. Regarding stability against displacements in $x$, we again obtain \textcolor{black}{the ``net pusher'' condition} $(S_{dd}^{(1)} + S_{dd}^{(2)}) < 0$. \textcolor{black}{From Eq. \ref{eq:headtotaildist}, this} implies $U_{s}^{(1)} > U_{s}^{(2)}$. The other stability conditions are $S_{dd}^{(2)} (1 + 3 \Gamma_1) + S_{dd}^{(1)} (1 + \Gamma_2) > 0$ and Eq. \textcolor{black}{48} in the \textcolor{black}{SM}. For axisymmetric swimmers, we can obtain stable pairing. Specifically, $\sigma_0^{(1)} +   \sigma_0^{(2)} + 3 (\sigma_0^{(2)} \Gamma_1 + \sigma_0^{(1)} \Gamma_2) > 0$ and Eq. 13 in the SI. Overall, the phase behavior is determined by four parameters: $\Gamma_1$, $\Gamma_2$, $S \equiv {\sigma_0^{(2)}}/{\sigma_0^{(1)}}$, and $V \equiv {U_{s}^{(2)}}/{U_{s}^{(1)}}$. For the slice of phase space in Fig. \ref{Fig:phase_head_tail} (a), we fix $V=0.8$ and $\Gamma_2=-0.8$, but vary $\Gamma_1$ and $S$. In the numerics, $B_1$ and $B_2$ are chosen to vary $S$ while keeping $V=0.8$. The model has good agreement with the numerics. There are two areas of significant disagreement. Similar to head-to-head pairs, one is for oblate particles with large $r_e$. The other is the slim area bordering $S=-1$, where $d_0 \rightarrow 0$. An example trajectory is shown in Fig. \ref{Fig:phase_head_tail} (b).
\begin{figure}[t]
\centering
\includegraphics[]{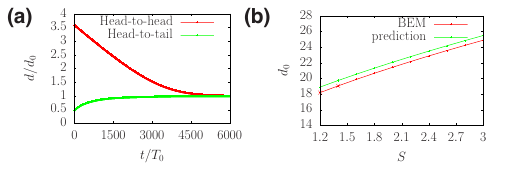}
\caption{(a) Separation $d$ for two squirmers forming a stable bound state. For the head-to-tail pair, $\Gamma_1=-0.7$, $\Gamma_2=-0.8$, $S=2.2$ and $U=0.8$, with $x_\text{2,initial}=3$ and $y_\text{2,initial}=10$. For the head-to-head pair, $B_1=0.1$, $B_2=-1$, $\tilde{B}=1.6833$ and $\Gamma=-0.835$, with $x_\text{initial}=2$ and $y_\text{initial}=55$. The particle parameters correspond to Figs. \ref{Fig:phase_head_head} (b) and \ref{Fig:phase_head_tail} (b), respectively. (b) The predicted and numerically calculated steady separations for head-to-tail pairs with $\Gamma_1=-0.7$, $\Gamma_2=-0.8$, $V=0.8$ and varying $S$, corresponding to the second row from the bottom in Fig. \ref{Fig:phase_head_tail} (a). Symbols indicate values of $S$ for which theory and numerics disagree concerning stability.}
\label{Fig:time_plot}
\end{figure}

\section{Conclusions} 
We have shown that non-spherical active particles can form bound pairs through far-field hydrodynamic interactions. A surprising finding of our work is that squirmers with non-axisymmetric surface slip may be capable of pairing behaviors that are \textit{not} obtainable for squirmers with axisymmetric slip. %These predictions could be tested using particles that move by induced charge electrophoresis. %The effective squirmer model can be an excellent approximation for the interaction of these particles, including in the presence of bounding surfaces \cite{katuri22}. %Here, we restricted our consideration to the DC limit, but the frequency of the driving AC field will also quantitatively impact the active stresslet and self-propulsion velocity, which can be addressed in future work.

We restricted our consideration to swimmers moving in the plane containing their center-to-center vector (the $xy$ plane). For non-axisymmetric particles defined by Eq. \ref{eq:delta}, a $90^{\circ}$ rotation of both particles around their $\hat{d}$ axes will invert the sign of $\delta$. Therefore, when head-to-head bound states, aligned with $x$, are stable against perturbations in the $xy$ plane, they will be unstable in the $yz$ plane. However, our quasi-2D assumption is realized in most active matter experiments. For head-to-tail pairs of axisymmetric particles, the stability conditions found here apply to general three-dimensional motions. 
 
Future work could incorporate the effects of inertia and/or near-field hydrodynamic interactions \cite{darveniza22,ouyang23}. Lubrication interactions can induce bound states for spherical squirmers near contact \cite{drescher09,darveniza22}.  \textcolor{black}{Additionally, making use of the Fax\'en relations for spheroids would account for the finite size of a particle in its response to ambient flow \cite{kim2013microhydrodynamics, claeys1993suspensions}.}  Our model may  have stable bound states in which particle orientations are not aligned in the direction of propulsion. Finally, the bound states found here may have implications for hierarchical self-organization and collective behavior. \textcolor{black}{For instance, Ref. \citenum{katuri22} observed that initial formation of immotile head-to-head bound states locally promoted  formation of additional bound states in a feedback loop, ultimately leading to phase separation. This  mechanism could be studied in the framework of the present work.} 
% Finally, we restricted our attention to bound states in which the particles' orientations are aligned with the center-to-center vector. This could be relaxed, potentially leading to bound states with circular trajectories.

\section{Acknowledgments}
We gratefully acknowledge donors of the American Chemical Society Petroleum Research Fund for support of this research through Grant No. 60809-DNI9.  \textcolor{black}{This research was also sponsored by the Army Research Office and was accomplished under Grant Number W911NF-23-1-0190. The views and conclusions contained in this document are those of the authors and should not be interpreted as representing the official policies, either expressed or implied, of the Army Research Office or the U.S. Government. The U.S. Government is authorized to reproduce and distribute reprints for Government purposes notwithstanding any copyright notation herein.} The technical support and advanced computing resources from University of Hawaii Information Technology Services – Cyberinfrastructure, funded in part by the National Science Foundation CC* awards \#2201428 and \#2232862 are gratefully acknowledged. We also thank Rumen Georgiev for insightful discussions.

\bibliography{spheroidal_squirmers}

\begin{thebibliography}{10}

\bibitem{FocalTverskyLossAbrahamISBI2019}
Nabila Abraham and Naimul~Mefraz Khan.
\newblock {A Novel Focal Tversky Loss Function with Improved Attention U-Net
  for Lesion Segmentation}.
\newblock {\em ISBI}, 2019.

\bibitem{Lovasz-softmaxLossBermanCVPR2018}
Maxim Berman, Amal~Rannen Triki, and Matthew~B. Blaschko.
\newblock {The Lovasz-softmax loss: A Tractable Surrogate for the Optimization
  of the Intersection-Over-Union Measure in Neural Networks}.
\newblock {\em CVPR}, 2018.

\bibitem{TheoreticalBertelsMIA2021}
Jeroen Bertels, David Robben, Dirk Vandermeulen, and Paul Suetens.
\newblock {Theoretical analysis and experimental validation of volume bias of
  soft Dice optimized segmentation maps in the context of inherent
  uncertainty}.
\newblock {\em MIA}, 2021.

\bibitem{MixMatchBerthelotNeurIPS2019}
David Berthelot, Nicholas Carlini, Ian Goodfellow, Nicolas Papernot, Avital
  Oliver, and Colin Raffel.
\newblock {MixMatch: A Holistic Approach to Semi-Supervised Learning}.
\newblock {\em NeurIPS}, 2019.

\bibitem{DETRCarionECCV2020}
Nicolas Carion, Francisco Massa, Gabriel Synnaeve, Nicolas Usunier, Alexander
  Kirillov, and Sergey Zagoruyko.
\newblock {End-to-End Object Detection with Transformers}.
\newblock {\em ECCV}, 2020.

\bibitem{DeepLabV3ChenarXiv2017}
Liang-Chieh Chen, George Papandreou, Florian Schroff, and Hartwig Adam.
\newblock {Rethinking Atrous convolution for semantic image segmentation}.
\newblock {\em arXiv}, 2017.

\bibitem{DeepLabV3+ChenECCV2018}
Liang-Chieh Chen, Yukun Zhu, George Papandreou, Florian Schroff, and Hartwig
  Adam.
\newblock {Encoder-decoder with Atrous separable convolution for semantic image
  segmentation}.
\newblock {\em ECCV}, 2018.

\bibitem{Mask2FormerChengCVPR2022}
Bowen Cheng, Ishan Misra, Alexander~G. Schwing, Alexander Kirillov, and Rohit
  Girdhar.
\newblock {Masked-attention Mask Transformer for Universal Image Segmentation}.
\newblock {\em CVPR}, 2022.

\bibitem{MaskFormerChengNeurIPS2021}
Bowen Cheng, Alexander~G Schwing, and Alexander Kirillov.
\newblock {Per-Pixel Classification is Not All You Need for Semantic
  Segmentation}.
\newblock {\em NeurIPS}, 2021.

\bibitem{MMSegmentation2020}
MMSegmentation Contributors.
\newblock {MMSegmentation: OpenMMLab Semantic Segmentation Toolbox and
  Benchmark}, 2020.

\bibitem{CityscapesCordtsCVPR2016}
Marius Cordts, Mohamed Omran, Sebastian Ramos, Timo Rehfeld, Markus Enzweiler,
  Rodrigo Benenson, Uwe Franke, Stefan Roth, and Bernt Schiele.
\newblock {The Cityscapes Dataset for Semantic Urban Scene Understanding}.
\newblock {\em CVPR}, 2016.

\bibitem{DeepGlobeDemirCVPRWorkshop2018}
Ilke Demir, Krzysztof Koperski, David Lindenbaum, Guan Pang, Jing Huang, Saikat
  Basu, Forest Hughes, Devis Tuia, and Ramesh Raskar.
\newblock {DeepGlobe 2018: A Challenge to Parse the Earth through Satellite
  Images}.
\newblock {\em CVPR Workshop}, 2018.

\bibitem{ImageNetDengCVPR2009}
Jia Deng, Wei Dong, Richard Socher, Li-Jia Li, Kai Li, and Li~Fei-Fei.
\newblock {ImageNet: A Large-Scale Hierarchical Image Database}.
\newblock {\em CVPR}, 2009.

\bibitem{EncyclopediaDeza2009}
Michel~Marie Deza and Elena Deza.
\newblock {\em {Encyclopedia of Distances}}.
\newblock Springer, 2009.

\bibitem{LTSDingICCV2021}
Zhipeng Ding, Xu~Han, Peirong Liu, and Marc Niethammer.
\newblock {Local Temperature Scaling for Probability Calibration}.
\newblock {\em ICCV}, 2021.

\bibitem{jSTABLDorentMIA2020}
Reuben Dorent, Thomas Booth, Wenqi Li, Carole~H. Sudre, Sina Kafiabadi, Jorge
  Cardoso, Sebastien Ourselin, and Tom Vercauteren.
\newblock {Learning joint segmentation of tissues and brain lesions from
  task-specific hetero-modal domain-shifted datasets}.
\newblock {\em MIA}, 2020.

\bibitem{OptimizationEelbodeTMI2020}
Tom Eelbode, Jeroen Bertels, Maxim Berman, Dirk Vandermeulen, Frederik Maes,
  Raf Bisschops, and Matthew~B. Blaschko.
\newblock {Optimization for medical image segmentation: Theory and Practice
  When Evaluating With Dice Score or Jaccard Index}.
\newblock {\em TMI}, 2020.

\bibitem{PASCALVOCEveringhamIJCV2009}
Mark Everingham, Luc~Van Gool, Christopher K~I Williams, John Winn, and Andrew
  Zisserman.
\newblock {The Pascal Visual Object Classes (VOC) Challenge}.
\newblock {\em IJCV}, 2009.

\bibitem{JIoUFengTITS2022}
Di~Feng, Zining Wang, Yiyang Zhou, Lars Rosenbaum, Fabian Timm, Klaus
  Dietmayer, Masayoshi Tomizuka, and Wei Zhan.
\newblock {Labels are Not Perfect: Inferring Spatial Uncertainty in Object
  Detection}.
\newblock {\em TITS}, 2022.

\bibitem{CalibrationGuoICML2017}
Chuan Guo, Geoff Pleiss, Yu~Sun, and Kilian~Q Weinberger.
\newblock {On calibration of modern neural networks}.
\newblock {\em ICML}, 2017.

\bibitem{ResNetHeCVPR2016}
Kaiming He, Xiangyu Zhang, Shaoqing Ren, and Jian Sun.
\newblock {Deep residual learning for image recognition}.
\newblock {\em CVPR}, 2016.

\bibitem{KDHintonNeurIPSWorkshop2015}
Geoffrey Hinton, Oriol Vinyals, and Jeff Dean.
\newblock {Distilling the knowledge in a neural network}.
\newblock {\em NeurIPS Workshop}, 2015.

\bibitem{DISTHuangNeurIPS2022}
Tao Huang, Shan You, Fei Wang, Chen Qian, and Chang Xu.
\newblock {Knowledge Distillation from A Stronger Teacher}.
\newblock {\em NeurIPS}, 2022.

\bibitem{MasKDHuangICLR2023}
Tao Huang, Yuan Zhang, Shan You, Fei Wang, Chen Qian, Jian Cao, and Chang Xu.
\newblock {Masked Distillation with Receptive Tokens}.
\newblock {\em ICLR}, 2023.

\bibitem{SMPIakubovskii2019}
Pavel Iakubovskii.
\newblock {Segmentation models pytorch}, 2019.

\bibitem{ImprovedIoffeICDM2010}
Sergey Ioffe.
\newblock {Improved Consistent Sampling, Weighted Minhash and L1 Sketching}.
\newblock {\em ICDM}, 2010.

\bibitem{nnU-NetIsenseeNatureMethods2021}
Fabian Isensee, Paul~F. Jaeger, Simon A.~A. Kohl, Jens Petersen, and Klaus~H.
  Maier-Hein.
\newblock {nnU-Net: a self-configuring method for deep learning-based
  biomedical image segmentation}.
\newblock {\em Nature Methods}, 2021.

\bibitem{PanopticSegmentationKirillovCVPR2019}
Alexander Kirillov, Kaiming He, Ross Girshick, Carsten Rother, and Piotr
  Dollar.
\newblock {Panoptic Segmentation}.
\newblock {\em CVPR}, 2019.

\bibitem{SAMKirillovICCV2023}
Alexander Kirillov, Eric Mintun, Nikhila Ravi, Hanzi Mao, Chloe Rolland, Laura
  Gustafson, Tete Xiao, Spencer Whitehead, Alexander~C. Berg, Wan-Yen Lo, Piotr
  Dollar, and Ross Girshick.
\newblock {Segment Anything}.
\newblock {\em ICCV}, 2023.

\bibitem{ANoteKosubPRL2019}
Sven Kosub.
\newblock {A note on the triangle inequality for the Jaccard distance}.
\newblock {\em PRL}, 2019.

\bibitem{MMCEKumarICML2018}
Aviral Kumar, Sunita Sarawagi, and Ujjwal Jain.
\newblock {Trainable Calibration Measures For Neural Networks From Kernel Mean
  Embeddings}.
\newblock {\em ICML}, 2018.

\bibitem{AutoLoss-ZeroLiCVPR2022}
Hao Li, Tianwen Fu, Jifeng Dai, Hongsheng Li, Gao Huang, and Xizhou Zhu.
\newblock {AutoLoss-Zero: Searching Loss Functions from Scratch for Generic
  Tasks}.
\newblock {\em CVPR}, 2022.

\bibitem{AutoSegLossLiICLR2021}
Hao Li, Chenxin Tao, Xizhou Zhu, Xiaogang Wang, Gao Huang, and Jifeng Dai.
\newblock {Auto Seg-Loss: Searching Metric Surrogates for Semantic
  Segmentation}.
\newblock {\em ICLR}, 2021.

\bibitem{DiceLossLiACL2020}
Xiaoya Li, Xiaofei Sun, Yuxian Meng, Junjun Liang, Fei Wu, and Jiwei Li.
\newblock {Dice Loss for Data-imbalanced NLP Tasks}.
\newblock {\em ACL}, 2020.

\bibitem{FocalLossLinTPAMI2018}
Tsung-Yi Lin, Priya Goyal, Ross Girshick, Kaiming He, and Piotr Dollár.
\newblock {Focal Loss for Dense Object Detection}.
\newblock {\em TPAMI}, 2018.

\bibitem{SKDLiuCVPR2019}
Yifan Liu, Ke~Chen, Chris Liu, Zengchang Qin, Zhenbo Luo, and Jingdong Wang.
\newblock {Structured Knowledge Distillation for Semantic Segmentation}.
\newblock {\em CVPR}, 2019.

\bibitem{ConvNeXtLiuCVPR2022}
Zhuang Liu, Hanzi Mao, Chao-Yuan Wu, Christoph Feichtenhofer, Trevor Darrell,
  and Saining Xie.
\newblock {A ConvNet for the 2020s}.
\newblock {\em CVPR}, 2022.

\bibitem{AdamWLoshchilovICLR2019}
Ilya Loshchilov and Frank Hutter.
\newblock {Decoupled Weight Decay Regularization}.
\newblock {\em ICLR}, 2019.

\bibitem{Meta-CalMaICML2021}
Xingchen Ma and Matthew~B Blaschko.
\newblock {Meta-Cal: Well-controlled Post-hoc Calibration by Ranking}.
\newblock {\em ICML}, 2021.

\bibitem{MetricsMaier-HeinarXiv2023}
Lena Maier-Hein et~al.
\newblock {Metrics Reloaded: Recommendations for image analysis validation}.
\newblock {\em arXiv}, 2023.

\bibitem{ConfidenceMehrtashTMI2020}
Alireza Mehrtash, William~M. Wells, Clare~M. Tempany, Purang Abolmaesumi, and
  Tina Kapur.
\newblock {Confidence Calibration and Predictive Uncertainty Estimation for
  Deep Medical Image Segmentation}.
\newblock {\em TMI}, 2020.

\bibitem{Self-DistillationMobahiNeurIPS2020}
Hossein Mobahi, Mehrdad Farajtabar, and Peter~L Bartlett.
\newblock {Self-Distillation Amplifies Regularization in Hilbert Space}.
\newblock {\em NeurIPS}, 2020.

\bibitem{MaximallyMoultonICDMWorkshop2018}
Ryan Moulton and Yunjiang Jiang.
\newblock {Maximally Consistent Sampling and the Jaccard Index of Probability
  Distributions}.
\newblock {\em ICDM Workshop}, 2018.

\bibitem{WhenMullerNeurIPS2019}
Rafael Müller, Simon Kornblith, and Geoffrey Hinton.
\newblock {When Does Label Smoothing Help?}
\newblock {\em NeurIPS}, 2019.

\bibitem{MeasuringNixonCVPRWorkshop2019}
Jeremy Nixon, Mike Dusenberry, Ghassen Jerfel, Timothy Nguyen, Jeremiah Liu,
  Linchuan Zhang, and Dustin Tran.
\newblock {Measuring Calibration in Deep Learning}.
\newblock {\em CVPR Workshop}, 2019.

\bibitem{OptimalNowozinCVPR2014}
Sebastian Nowozin.
\newblock {Optimal Decisions from Probabilistic Models: the
  Intersection-over-Union Case}.
\newblock {\em CVPR}, 2014.

\bibitem{OnPopordanoskaMICCAI2021}
Teodora Popordanoska, Jeroen Bertels, Dirk Vandermeulen, Frederik Maes, and
  Matthew~B. Blaschko.
\newblock {On the relationship between calibrated predictors and unbiased
  volume estimation}.
\newblock {\em MICCAI}, 2021.

\bibitem{KDE-XEPopordanoskaNeurIPS2022}
Teodora Popordanoska, Raphael Sayer, and Matthew~B Blaschko.
\newblock {A Consistent and Differentiable Lp Canonical Calibration Error
  Estimator}.
\newblock {\em NeurIPS}, 2022.

\bibitem{SoftJaccardRahmanISVC2016}
Md~Atiqur Rahman and Yang Wang.
\newblock {Optimizing intersection-over-union in deep neural networks for image
  segmentation}.
\newblock {\em ISVC}, 2016.

\bibitem{LandRakhlinCVPRWorkshop2018}
Alexander Rakhlin, Alex Davydow, and Sergey Nikolenko.
\newblock {Land Cover Classification from Satellite Imagery With U-Net and
  Lovász-Softmax Loss}.
\newblock {\em CVPR Workshop}, 2018.

\bibitem{U-NetRonnebergerMICCAI2015}
Olaf Ronneberger, Philipp Fischer, and Thomas Brox.
\newblock {U-Net: Convolutional Networks for Biomedical Image Segmentation}.
\newblock {\em MICCAI}, 2015.

\bibitem{PostRousseauISBI2021}
Axel-Jan Rousseau, Thijs Becker, Jeroen Bertels, Matthew~B. Blaschko, and Dirk
  Valkenborg.
\newblock {Post training uncertainty calibration of deep networks for medical
  image segmentation}.
\newblock {\em ISBI}, 2021.

\bibitem{SoftTverskyLossSalehiMICCAIWorkshop2017}
Seyed Sadegh~Mohseni Salehi, Deniz Erdogmus, and Ali Gholipour.
\newblock {Tversky loss function for image segmentation using 3D fully
  convolutional deep networks}.
\newblock {\em MICCAI Workshop}, 2017.

\bibitem{MobileNetV2SandlerCVPR2018}
Mark Sandler, Andrew Howard, Menglong Zhu, Andrey Zhmoginov, and Liang-Chieh
  Chen.
\newblock {MobileNetV2: Inverted Residuals and Linear Bottlenecks}.
\newblock {\em CVPR}, 2018.

\bibitem{CDShuICCV2021}
Changyong Shu, Yifan Liu, Jianfei Gao, Zheng Yan, and Chunhua Shen.
\newblock {Channel-wise Knowledge Distillation for Dense Prediction}.
\newblock {\em ICCV}, 2021.

\bibitem{TheMinisumSpathOR1981}
H.~Späth.
\newblock {The minisum location problem for the Jaccard metric}.
\newblock {\em OR Spektrum}, 1981.

\bibitem{SoftDiceLossSudreMICCAIWorkshop2017}
Carole~H Sudre, Wenqi Li, Tom Vercauteren, Sébastien Ourselin, and M~Jorge
  Cardoso.
\newblock {Generalised Dice overlap as a deep learning loss function for highly
  unbalanced segmentations}.
\newblock {\em MICCAI Workshop}, 2017.

\bibitem{InceptionV2V3SzegedyCVPR2016}
Christian Szegedy, Vincent Vanhoucke, Sergey Ioffe, Jonathon Shlens, and
  Zbigniew Wojna.
\newblock {Rethinking the Inception Architecture for Computer Vision}.
\newblock {\em CVPR}, 2016.

\bibitem{UnderstandingTangarXiv2020}
Jiaxi Tang, Rakesh Shivanna, Zhe Zhao, Dong Lin, Anima Singh, Ed~H Chi, and
  Sagar Jain.
\newblock {Understanding and Improving Knowledge Distillation}.
\newblock {\em arXiv}, 2020.

\bibitem{NatureVapnik1995}
Vladimir~N Vapnik.
\newblock {\em {The Nature of Statistical Learning Theory}}.
\newblock Springer, 1995.

\bibitem{MaX-DeepLabWangCVPR2021}
Huiyu Wang, Yukun Zhu, Hartwig Adam, Alan Yuille, and Liang-Chieh Chen.
\newblock {MaX-DeepLab: End-to-End Panoptic Segmentation with Mask
  Transformers}.
\newblock {\em CVPR}, 2021.

\bibitem{InternImageWangCVPR2023}
Wenhai Wang, Jifeng Dai, Zhe Chen, Zhenhang Huang, Zhiqi Li, Xizhou Zhu,
  Xiaowei Hu, Tong Lu, Lewei Lu, Hongsheng Li, Xiaogang Wang, and Yu~Qiao.
\newblock {InternImage: Exploring Large-Scale Vision Foundation Models with
  Deformable Convolutions}.
\newblock {\em CVPR}, 2023.

\bibitem{U2PLWangCVPR2022}
Yuchao Wang, Haochen Wang, Yujun Shen, Jingjing Fei, Wei Li, Guoqiang Jin,
  Liwei Wu, Rui Zhao, and Xinyi Le.
\newblock {Semi-Supervised Semantic Segmentation Using Unreliable
  Pseudo-Labels}.
\newblock {\em CVPR}, 2022.

\bibitem{IFVDWangECCV2020}
Yukang Wang, Wei Zhou, Tao Jiang, Xiang Bai, and Yongchao Xu.
\newblock {Intra-class Feature Variation Distillation for Semantic
  Segmentation}.
\newblock {\em ECCV}, 2020.

\bibitem{Fine-grainedIoUsWangNeurIPS2023}
Zifu Wang, Maxim Berman, Amal Rannen-Triki, Philip~H.S. Torr, Devis Tuia, Tinne
  Tuytelaars, Luc~Van Gool, Jiaqian Yu, and Matthew~B. Blaschko.
\newblock {Revisiting Evaluation Metrics for Semantic Segmentation:
  Optimization and Evaluation of Fine-grained Intersection over Union}.
\newblock {\em NeurIPS}, 2023.

\bibitem{DMLWangMICCAI2023}
Zifu Wang, Teodora Popordanoska, Jeroen Bertels, Robin Lemmens, and Matthew~B.
  Blaschko.
\newblock {Dice Semimetric Losses: Optimizing the Dice Score with Soft Labels}.
\newblock {\em MICCAI}, 2023.

\bibitem{timmWightman2019}
Ross Wightman.
\newblock {Pytorch Image Models}, 2019.

\bibitem{ConvNeXtV2WooCVPR2023}
Sanghyun Woo, Shoubhik Debnath, Ronghang Hu, Xinlei Chen, Zhuang Liu, In~So
  Kweon, and Saining Xie.
\newblock {ConvNeXt V2: Co-designing and Scaling ConvNets with Masked
  Autoencoders}.
\newblock {\em CVPR}, 2023.

\bibitem{SegFormerXieNeurIPS2021}
Enze Xie, Wenhai Wang, Zhiding Yu, Anima Anandkumar, Jose~M Alvarez, and Ping
  Luo.
\newblock {SegFormer: Simple and Efficient Design for Semantic Segmentation
  with Transformers}.
\newblock {\em NeurIPS}, 2021.

\bibitem{CIRKDYangCVPR2022}
Chuanguang Yang, Helong Zhou, Zhulin An, Xue Jiang, Yongjun Xu, and Qian Zhang.
\newblock {Cross-Image Relational Knowledge Distillation for Semantic
  Segmentation}.
\newblock {\em CVPR}, 2022.

\bibitem{UniMatchYangCVPR2023}
Lihe Yang, Lei Qi, Litong Feng, Wayne Zhang, and Yinghuan Shi.
\newblock {Revisiting Weak-to-Strong Consistency in Semi-Supervised Semantic
  Segmentation}.
\newblock {\em CVPR}, 2023.

\bibitem{LovaszHingeYuTPAMI2018}
Jiaqian Yu and Matthew~B. Blaschko.
\newblock {The Lovász Hinge: A Novel Convex Surrogate for Submodular Losses}.
\newblock {\em TPAMI}, 2018.

\bibitem{PixIoUYuICML2021}
Jiaqian Yu, Jingtao Xu, Yiwei Chen, Weiming Li, Qiang Wang, Byung~In Yoo, and
  Jae-Joon Han.
\newblock {Learning generalized intersection over union for dense pixelwise
  prediction}.
\newblock {\em ICML}, 2021.

\bibitem{CMT-DeepLabYuCVPR2022}
Qihang Yu, Huiyu Wang, Dahun Kim, Siyuan Qiao, Maxwell Collins, Yukun Zhu,
  Hartwig Adam, Alan Yuille, and Liang-Chieh Chen.
\newblock {CMT-DeepLab: Clustering Mask Transformers for Panoptic
  Segmentation}.
\newblock {\em CVPR}, 2022.

\bibitem{kMaX-DeepLabYuECCV2022}
Qihang Yu, Huiyu Wang, Siyuan Qiao, Maxwell Collins, Yukun Zhu, Hatwig Adam,
  Alan Yuille, and Liang-Chieh Chen.
\newblock {k-means Mask Transformer}.
\newblock {\em ECCV}, 2022.

\bibitem{Tf-KDYuanCVPR2020}
Li~Yuan, Francis~EH Tay, Guilin Li, Tao Wang, and Jiashi Feng.
\newblock {Revisiting Knowledge Distillation via Label Smoothing
  Regularization}.
\newblock {\em CVPR}, 2020.

\bibitem{DKDZhaoCVPR2022}
Borui Zhao, Quan Cui, Renjie Song, Yiyu Qiu, and Jiajun Liang.
\newblock {Decoupled Knowledge Distillation}.
\newblock {\em CVPR}, 2022.

\bibitem{PSPNetZhaoCVPR2017}
Hengshuang Zhao, Jianping Shi, Xiaojuan Qi, Xiaogang Wang, and Jiaya Jia.
\newblock {Pyramid Scene Parsing Network}.
\newblock {\em CVPR}, 2017.

\bibitem{iMASZhaoCVPR2023}
Zhen Zhao, Sifan Long, Jimin Pi, Jingdong Wang, and Luping Zhou.
\newblock {Instance-specific and Model-adaptive Supervision for Semi-supervised
  Semantic Segmentation}.
\newblock {\em CVPR}, 2023.

\bibitem{AugSegZhaoCVPR2023}
Zhen Zhao, Lihe Yang, Sifan Long, Jimin Pi, Luping Zhou, and Jingdong Wang.
\newblock {Augmentation Matters: A Simple-yet-Effective Approach to
  Semi-supervised Semantic Segmentation}.
\newblock {\em CVPR}, 2023.

\bibitem{ADE20KZhouCVPR2017}
Bolei Zhou, Hang Zhao, Xavier Puig, Sanja Fidler, Adela Barriuso, and Antonio
  Torralba.
\newblock {Scene Parsing Through ADE20K Dataset}.
\newblock {\em CVPR}, 2017.

\bibitem{RethinkingZhouICLR2021}
Helong Zhou, Liangchen Song, Jiajie Chen, Ye~Zhou, Guoli Wang, Junsong Yuan,
  and Qian Zhang.
\newblock {Rethinking Soft Labels for Knowledge Distillation: A Bias-Variance
  Tradeoff Perspective}.
\newblock {\em ICLR}, 2021.

\end{thebibliography}


\begin{thebibliography}{74}%
\makeatletter
\providecommand \@ifxundefined [1]{%
 \@ifx{#1\undefined}
}%
\providecommand \@ifnum [1]{%
 \ifnum #1\expandafter \@firstoftwo
 \else \expandafter \@secondoftwo
 \fi
}%
\providecommand \@ifx [1]{%
 \ifx #1\expandafter \@firstoftwo
 \else \expandafter \@secondoftwo
 \fi
}%
\providecommand \natexlab [1]{#1}%
\providecommand \enquote  [1]{``#1''}%
\providecommand \bibnamefont  [1]{#1}%
\providecommand \bibfnamefont [1]{#1}%
\providecommand \citenamefont [1]{#1}%
\providecommand \href@noop [0]{\@secondoftwo}%
\providecommand \href [0]{\begingroup \@sanitize@url \@href}%
\providecommand \@href[1]{\@@startlink{#1}\@@href}%
\providecommand \@@href[1]{\endgroup#1\@@endlink}%
\providecommand \@sanitize@url [0]{\catcode `\\12\catcode `\$12\catcode
  `\&12\catcode `\#12\catcode `\^12\catcode `\_12\catcode `\%12\relax}%
\providecommand \@@startlink[1]{}%
\providecommand \@@endlink[0]{}%
\providecommand \url  [0]{\begingroup\@sanitize@url \@url }%
\providecommand \@url [1]{\endgroup\@href {#1}{\urlprefix }}%
\providecommand \urlprefix  [0]{URL }%
\providecommand \Eprint [0]{\href }%
\providecommand \doibase [0]{https://doi.org/}%
\providecommand \selectlanguage [0]{\@gobble}%
\providecommand \bibinfo  [0]{\@secondoftwo}%
\providecommand \bibfield  [0]{\@secondoftwo}%
\providecommand \translation [1]{[#1]}%
\providecommand \BibitemOpen [0]{}%
\providecommand \bibitemStop [0]{}%
\providecommand \bibitemNoStop [0]{.\EOS\space}%
\providecommand \EOS [0]{\spacefactor3000\relax}%
\providecommand \BibitemShut  [1]{\csname bibitem#1\endcsname}%
\let\auto@bib@innerbib\@empty
%</preamble>
\bibitem [{\citenamefont {Sanchez}\ \emph {et~al.}(2012)\citenamefont
  {Sanchez}, \citenamefont {Chen}, \citenamefont {DeCamp}, \citenamefont
  {Heymann},\ and\ \citenamefont {Dogic}}]{sanchez2012}%
  \BibitemOpen
  \bibfield  {author} {\bibinfo {author} {\bibfnamefont {T.}~\bibnamefont
  {Sanchez}}, \bibinfo {author} {\bibfnamefont {D.~T.}\ \bibnamefont {Chen}},
  \bibinfo {author} {\bibfnamefont {S.~J.}\ \bibnamefont {DeCamp}}, \bibinfo
  {author} {\bibfnamefont {M.}~\bibnamefont {Heymann}},\ and\ \bibinfo {author}
  {\bibfnamefont {Z.}~\bibnamefont {Dogic}},\ }\bibfield  {title} {\bibinfo
  {title} {Spontaneous motion in hierarchically assembled active matter},\
  }\href@noop {} {\bibfield  {journal} {\bibinfo  {journal} {Nature}\ }\textbf
  {\bibinfo {volume} {491}},\ \bibinfo {pages} {431} (\bibinfo {year}
  {2012})}\BibitemShut {NoStop}%
\bibitem [{\citenamefont {Aubret}\ \emph {et~al.}(2018)\citenamefont {Aubret},
  \citenamefont {Youssef}, \citenamefont {Sacanna},\ and\ \citenamefont
  {Palacci}}]{aubret2018}%
  \BibitemOpen
  \bibfield  {author} {\bibinfo {author} {\bibfnamefont {A.}~\bibnamefont
  {Aubret}}, \bibinfo {author} {\bibfnamefont {M.}~\bibnamefont {Youssef}},
  \bibinfo {author} {\bibfnamefont {S.}~\bibnamefont {Sacanna}},\ and\ \bibinfo
  {author} {\bibfnamefont {J.}~\bibnamefont {Palacci}},\ }\bibfield  {title}
  {\bibinfo {title} {Targeted assembly and synchronization of self-spinning
  microgears},\ }\href@noop {} {\bibfield  {journal} {\bibinfo  {journal}
  {Nature Physics}\ }\textbf {\bibinfo {volume} {14}},\ \bibinfo {pages} {1114}
  (\bibinfo {year} {2018})}\BibitemShut {NoStop}%
\bibitem [{\citenamefont {Boymelgreen}\ \emph {et~al.}(2018)\citenamefont
  {Boymelgreen}, \citenamefont {Balli}, \citenamefont {Miloh},\ and\
  \citenamefont {Yossifon}}]{boymelgreen2018}%
  \BibitemOpen
  \bibfield  {author} {\bibinfo {author} {\bibfnamefont {A.~M.}\ \bibnamefont
  {Boymelgreen}}, \bibinfo {author} {\bibfnamefont {T.}~\bibnamefont {Balli}},
  \bibinfo {author} {\bibfnamefont {T.}~\bibnamefont {Miloh}},\ and\ \bibinfo
  {author} {\bibfnamefont {G.}~\bibnamefont {Yossifon}},\ }\bibfield  {title}
  {\bibinfo {title} {Active colloids as mobile microelectrodes for unified
  label-free selective cargo transport},\ }\href@noop {} {\bibfield  {journal}
  {\bibinfo  {journal} {Nature communications}\ }\textbf {\bibinfo {volume}
  {9}},\ \bibinfo {pages} {760} (\bibinfo {year} {2018})}\BibitemShut {NoStop}%
\bibitem [{\citenamefont {Arora}\ \emph {et~al.}(2021)\citenamefont {Arora},
  \citenamefont {Sood},\ and\ \citenamefont {Ganapathy}}]{Arora21}%
  \BibitemOpen
  \bibfield  {author} {\bibinfo {author} {\bibfnamefont {P.}~\bibnamefont
  {Arora}}, \bibinfo {author} {\bibfnamefont {A.~K.}\ \bibnamefont {Sood}},\
  and\ \bibinfo {author} {\bibfnamefont {R.}~\bibnamefont {Ganapathy}},\
  }\bibfield  {title} {\bibinfo {title} {Emergent stereoselective interactions
  and self-recognition in polar chiral active ellipsoids},\ }\href@noop {}
  {\bibfield  {journal} {\bibinfo  {journal} {Science Advances}\ }\textbf
  {\bibinfo {volume} {7}},\ \bibinfo {pages} {eabd0331} (\bibinfo {year}
  {2021})}\BibitemShut {NoStop}%
\bibitem [{\citenamefont {Palacci}\ \emph {et~al.}(2013)\citenamefont
  {Palacci}, \citenamefont {Sacanna}, \citenamefont {Steinberg}, \citenamefont
  {Pine},\ and\ \citenamefont {Chaikin}}]{palacci2013}%
  \BibitemOpen
  \bibfield  {author} {\bibinfo {author} {\bibfnamefont {J.}~\bibnamefont
  {Palacci}}, \bibinfo {author} {\bibfnamefont {S.}~\bibnamefont {Sacanna}},
  \bibinfo {author} {\bibfnamefont {A.~P.}\ \bibnamefont {Steinberg}}, \bibinfo
  {author} {\bibfnamefont {D.~J.}\ \bibnamefont {Pine}},\ and\ \bibinfo
  {author} {\bibfnamefont {P.~M.}\ \bibnamefont {Chaikin}},\ }\bibfield
  {title} {\bibinfo {title} {Living crystals of light-activated colloidal
  surfers},\ }\href@noop {} {\bibfield  {journal} {\bibinfo  {journal}
  {Science}\ }\textbf {\bibinfo {volume} {339}},\ \bibinfo {pages} {936}
  (\bibinfo {year} {2013})}\BibitemShut {NoStop}%
\bibitem [{\citenamefont {Buttinoni}\ \emph {et~al.}(2013)\citenamefont
  {Buttinoni}, \citenamefont {Bialk\'e}, \citenamefont {K\"ummel},
  \citenamefont {L\"owen}, \citenamefont {Bechinger},\ and\ \citenamefont
  {Speck}}]{Buttinoni13}%
  \BibitemOpen
  \bibfield  {author} {\bibinfo {author} {\bibfnamefont {I.}~\bibnamefont
  {Buttinoni}}, \bibinfo {author} {\bibfnamefont {J.}~\bibnamefont {Bialk\'e}},
  \bibinfo {author} {\bibfnamefont {F.}~\bibnamefont {K\"ummel}}, \bibinfo
  {author} {\bibfnamefont {H.}~\bibnamefont {L\"owen}}, \bibinfo {author}
  {\bibfnamefont {C.}~\bibnamefont {Bechinger}},\ and\ \bibinfo {author}
  {\bibfnamefont {T.}~\bibnamefont {Speck}},\ }\bibfield  {title} {\bibinfo
  {title} {Dynamical clustering and phase separation in suspensions of
  self-propelled colloidal particles},\ }\href@noop {} {\bibfield  {journal}
  {\bibinfo  {journal} {Phys. Rev. Lett.}\ }\textbf {\bibinfo {volume} {110}},\
  \bibinfo {pages} {238301} (\bibinfo {year} {2013})}\BibitemShut {NoStop}%
\bibitem [{\citenamefont {Pohl}\ and\ \citenamefont {Stark}(2014)}]{pohl14}%
  \BibitemOpen
  \bibfield  {author} {\bibinfo {author} {\bibfnamefont {O.}~\bibnamefont
  {Pohl}}\ and\ \bibinfo {author} {\bibfnamefont {H.}~\bibnamefont {Stark}},\
  }\bibfield  {title} {\bibinfo {title} {Dynamic clustering and chemotactic
  collapse of self-phoretic active particles},\ }\href
  {https://doi.org/10.1103/PhysRevLett.112.238303} {\bibfield  {journal}
  {\bibinfo  {journal} {Phys. Rev. Lett.}\ }\textbf {\bibinfo {volume} {112}},\
  \bibinfo {pages} {238303} (\bibinfo {year} {2014})}\BibitemShut {NoStop}%
\bibitem [{\citenamefont {Cates}\ and\ \citenamefont
  {Tailleur}(2015)}]{cates2015}%
  \BibitemOpen
  \bibfield  {author} {\bibinfo {author} {\bibfnamefont {M.~E.}\ \bibnamefont
  {Cates}}\ and\ \bibinfo {author} {\bibfnamefont {J.}~\bibnamefont
  {Tailleur}},\ }\bibfield  {title} {\bibinfo {title} {Motility-induced phase
  separation},\ }\href@noop {} {\bibfield  {journal} {\bibinfo  {journal}
  {Annu. Rev. Condens. Matter Phys.}\ }\textbf {\bibinfo {volume} {6}},\
  \bibinfo {pages} {219} (\bibinfo {year} {2015})}\BibitemShut {NoStop}%
\bibitem [{\citenamefont {Hokmabad}\ \emph {et~al.}(2022)\citenamefont
  {Hokmabad}, \citenamefont {Nishide}, \citenamefont {Ramesh}, \citenamefont
  {Kr{\"u}ger},\ and\ \citenamefont {Maass}}]{hokmabad2022}%
  \BibitemOpen
  \bibfield  {author} {\bibinfo {author} {\bibfnamefont {B.~V.}\ \bibnamefont
  {Hokmabad}}, \bibinfo {author} {\bibfnamefont {A.}~\bibnamefont {Nishide}},
  \bibinfo {author} {\bibfnamefont {P.}~\bibnamefont {Ramesh}}, \bibinfo
  {author} {\bibfnamefont {C.}~\bibnamefont {Kr{\"u}ger}},\ and\ \bibinfo
  {author} {\bibfnamefont {C.~C.}\ \bibnamefont {Maass}},\ }\bibfield  {title}
  {\bibinfo {title} {Spontaneously rotating clusters of active droplets},\
  }\href@noop {} {\bibfield  {journal} {\bibinfo  {journal} {Soft Matter}\
  }\textbf {\bibinfo {volume} {18}},\ \bibinfo {pages} {2731} (\bibinfo {year}
  {2022})}\BibitemShut {NoStop}%
\bibitem [{\citenamefont {Bricard}\ \emph {et~al.}(2013)\citenamefont
  {Bricard}, \citenamefont {Caussin}, \citenamefont {Desreumaux}, \citenamefont
  {Dauchot},\ and\ \citenamefont {Bartolo}}]{bricard13}%
  \BibitemOpen
  \bibfield  {author} {\bibinfo {author} {\bibfnamefont {A.}~\bibnamefont
  {Bricard}}, \bibinfo {author} {\bibfnamefont {J.}~\bibnamefont {Caussin}},
  \bibinfo {author} {\bibfnamefont {N.}~\bibnamefont {Desreumaux}}, \bibinfo
  {author} {\bibfnamefont {O.}~\bibnamefont {Dauchot}},\ and\ \bibinfo {author}
  {\bibfnamefont {D.}~\bibnamefont {Bartolo}},\ }\bibfield  {title} {\bibinfo
  {title} {Emergence of macroscopic directed motion in populations of motile
  colloids},\ }\href@noop {} {\bibfield  {journal} {\bibinfo  {journal}
  {Nature}\ }\textbf {\bibinfo {volume} {503}},\ \bibinfo {pages} {95}
  (\bibinfo {year} {2013})}\BibitemShut {NoStop}%
\bibitem [{\citenamefont {Yan}\ \emph {et~al.}(2016)\citenamefont {Yan},
  \citenamefont {Han}, \citenamefont {Zhang}, \citenamefont {Xu}, \citenamefont
  {Luijten},\ and\ \citenamefont {Granick}}]{yan2016}%
  \BibitemOpen
  \bibfield  {author} {\bibinfo {author} {\bibfnamefont {J.}~\bibnamefont
  {Yan}}, \bibinfo {author} {\bibfnamefont {M.}~\bibnamefont {Han}}, \bibinfo
  {author} {\bibfnamefont {J.}~\bibnamefont {Zhang}}, \bibinfo {author}
  {\bibfnamefont {C.}~\bibnamefont {Xu}}, \bibinfo {author} {\bibfnamefont
  {E.}~\bibnamefont {Luijten}},\ and\ \bibinfo {author} {\bibfnamefont
  {S.}~\bibnamefont {Granick}},\ }\bibfield  {title} {\bibinfo {title}
  {Reconfiguring active particles by electrostatic imbalance},\ }\href@noop {}
  {\bibfield  {journal} {\bibinfo  {journal} {Nature materials}\ }\textbf
  {\bibinfo {volume} {15}},\ \bibinfo {pages} {1095} (\bibinfo {year}
  {2016})}\BibitemShut {NoStop}%
\bibitem [{\citenamefont {Kaiser}\ \emph {et~al.}(2017)\citenamefont {Kaiser},
  \citenamefont {Snezhko},\ and\ \citenamefont {Aranson}}]{kaiser2017flocking}%
  \BibitemOpen
  \bibfield  {author} {\bibinfo {author} {\bibfnamefont {A.}~\bibnamefont
  {Kaiser}}, \bibinfo {author} {\bibfnamefont {A.}~\bibnamefont {Snezhko}},\
  and\ \bibinfo {author} {\bibfnamefont {I.~S.}\ \bibnamefont {Aranson}},\
  }\bibfield  {title} {\bibinfo {title} {Flocking ferromagnetic colloids},\
  }\href@noop {} {\bibfield  {journal} {\bibinfo  {journal} {Science Advances}\
  }\textbf {\bibinfo {volume} {3}},\ \bibinfo {pages} {e1601469} (\bibinfo
  {year} {2017})}\BibitemShut {NoStop}%
\bibitem [{\citenamefont {Han}\ \emph {et~al.}(2020)\citenamefont {Han},
  \citenamefont {Kokot}, \citenamefont {Tovkach}, \citenamefont {Glatz},
  \citenamefont {Aranson},\ and\ \citenamefont {Snezhko}}]{han2020}%
  \BibitemOpen
  \bibfield  {author} {\bibinfo {author} {\bibfnamefont {K.}~\bibnamefont
  {Han}}, \bibinfo {author} {\bibfnamefont {G.}~\bibnamefont {Kokot}}, \bibinfo
  {author} {\bibfnamefont {O.}~\bibnamefont {Tovkach}}, \bibinfo {author}
  {\bibfnamefont {A.}~\bibnamefont {Glatz}}, \bibinfo {author} {\bibfnamefont
  {I.~S.}\ \bibnamefont {Aranson}},\ and\ \bibinfo {author} {\bibfnamefont
  {A.}~\bibnamefont {Snezhko}},\ }\bibfield  {title} {\bibinfo {title}
  {Emergence of self-organized multivortex states in flocks of active
  rollers},\ }\href@noop {} {\bibfield  {journal} {\bibinfo  {journal}
  {Proceedings of the National Academy of Sciences}\ }\textbf {\bibinfo
  {volume} {117}},\ \bibinfo {pages} {9706} (\bibinfo {year}
  {2020})}\BibitemShut {NoStop}%
\bibitem [{\citenamefont {Zhang}\ \emph {et~al.}(2021)\citenamefont {Zhang},
  \citenamefont {Alert}, \citenamefont {Yan}, \citenamefont {Wingreen},\ and\
  \citenamefont {Granick}}]{Zhang21}%
  \BibitemOpen
  \bibfield  {author} {\bibinfo {author} {\bibfnamefont {J.}~\bibnamefont
  {Zhang}}, \bibinfo {author} {\bibfnamefont {R.}~\bibnamefont {Alert}},
  \bibinfo {author} {\bibfnamefont {J.}~\bibnamefont {Yan}}, \bibinfo {author}
  {\bibfnamefont {N.}~\bibnamefont {Wingreen}},\ and\ \bibinfo {author}
  {\bibfnamefont {S.}~\bibnamefont {Granick}},\ }\bibfield  {title} {\bibinfo
  {title} {Active phase separation by turning towards regions of higher
  density},\ }\href@noop {} {\bibfield  {journal} {\bibinfo  {journal} {Nat.
  Phys.}\ }\textbf {\bibinfo {volume} {17}} (\bibinfo {year}
  {2021})}\BibitemShut {NoStop}%
\bibitem [{\citenamefont {Mart{\'\i}nez-Pedrero}\ and\ \citenamefont
  {Tierno}(2018)}]{martinez2018advances}%
  \BibitemOpen
  \bibfield  {author} {\bibinfo {author} {\bibfnamefont {F.}~\bibnamefont
  {Mart{\'\i}nez-Pedrero}}\ and\ \bibinfo {author} {\bibfnamefont
  {P.}~\bibnamefont {Tierno}},\ }\bibfield  {title} {\bibinfo {title} {Advances
  in colloidal manipulation and transport via hydrodynamic interactions},\
  }\href@noop {} {\bibfield  {journal} {\bibinfo  {journal} {Journal of Colloid
  and Interface Science}\ }\textbf {\bibinfo {volume} {519}},\ \bibinfo {pages}
  {296} (\bibinfo {year} {2018})}\BibitemShut {NoStop}%
\bibitem [{\citenamefont {Lighthill}(1952)}]{Lighthill52}%
  \BibitemOpen
  \bibfield  {author} {\bibinfo {author} {\bibfnamefont {M.~J.}\ \bibnamefont
  {Lighthill}},\ }\bibfield  {title} {\bibinfo {title} {On the squirming motion
  of nearly spherical deformable bodies through liquids at very small reynolds
  numbers},\ }\href@noop {} {\bibfield  {journal} {\bibinfo  {journal} {Commun.
  Pure Appl. Math.}\ }\textbf {\bibinfo {volume} {5}},\ \bibinfo {pages} {109}
  (\bibinfo {year} {1952})}\BibitemShut {NoStop}%
\bibitem [{\citenamefont {Blake}(1971)}]{Blake71}%
  \BibitemOpen
  \bibfield  {author} {\bibinfo {author} {\bibfnamefont {J.~R.}\ \bibnamefont
  {Blake}},\ }\bibfield  {title} {\bibinfo {title} {A spherical envelope
  approach to ciliary propulsion},\ }\href@noop {} {\bibfield  {journal}
  {\bibinfo  {journal} {J. Fluid Mech.}\ }\textbf {\bibinfo {volume} {46}},\
  \bibinfo {pages} {199} (\bibinfo {year} {1971})}\BibitemShut {NoStop}%
\bibitem [{\citenamefont {Pedley}(2016)}]{pedley2016spherical}%
  \BibitemOpen
  \bibfield  {author} {\bibinfo {author} {\bibfnamefont {T.~J.}\ \bibnamefont
  {Pedley}},\ }\bibfield  {title} {\bibinfo {title} {Spherical squirmers:
  models for swimming micro-organisms},\ }\href@noop {} {\bibfield  {journal}
  {\bibinfo  {journal} {IMA Journal of Applied Mathematics}\ }\textbf {\bibinfo
  {volume} {81}},\ \bibinfo {pages} {488} (\bibinfo {year} {2016})}\BibitemShut
  {NoStop}%
\bibitem [{\citenamefont {Ishikawa}\ \emph {et~al.}(2006)\citenamefont
  {Ishikawa}, \citenamefont {Simmonds},\ and\ \citenamefont
  {Pedley}}]{ishikawa2006}%
  \BibitemOpen
  \bibfield  {author} {\bibinfo {author} {\bibfnamefont {T.}~\bibnamefont
  {Ishikawa}}, \bibinfo {author} {\bibfnamefont {M.}~\bibnamefont {Simmonds}},\
  and\ \bibinfo {author} {\bibfnamefont {T.~J.}\ \bibnamefont {Pedley}},\
  }\bibfield  {title} {\bibinfo {title} {Hydrodynamic interaction of two
  swimming model micro-organisms},\ }\href@noop {} {\bibfield  {journal}
  {\bibinfo  {journal} {Journal of Fluid Mechanics}\ }\textbf {\bibinfo
  {volume} {568}},\ \bibinfo {pages} {119} (\bibinfo {year}
  {2006})}\BibitemShut {NoStop}%
\bibitem [{\citenamefont {Ishikawa}\ and\ \citenamefont
  {Pedley}(2007)}]{ishikawa2007diffusion}%
  \BibitemOpen
  \bibfield  {author} {\bibinfo {author} {\bibfnamefont {T.}~\bibnamefont
  {Ishikawa}}\ and\ \bibinfo {author} {\bibfnamefont {T.}~\bibnamefont
  {Pedley}},\ }\bibfield  {title} {\bibinfo {title} {Diffusion of swimming
  model micro-organisms in a semi-dilute suspension},\ }\href@noop {}
  {\bibfield  {journal} {\bibinfo  {journal} {Journal of Fluid Mechanics}\
  }\textbf {\bibinfo {volume} {588}},\ \bibinfo {pages} {437} (\bibinfo {year}
  {2007})}\BibitemShut {NoStop}%
\bibitem [{\citenamefont {Drescher}\ \emph {et~al.}(2009)\citenamefont
  {Drescher}, \citenamefont {Leptos}, \citenamefont {Tuval}, \citenamefont
  {Ishikawa}, \citenamefont {Pedley},\ and\ \citenamefont
  {Goldstein}}]{drescher09}%
  \BibitemOpen
  \bibfield  {author} {\bibinfo {author} {\bibfnamefont {K.}~\bibnamefont
  {Drescher}}, \bibinfo {author} {\bibfnamefont {K.~C.}\ \bibnamefont
  {Leptos}}, \bibinfo {author} {\bibfnamefont {I.}~\bibnamefont {Tuval}},
  \bibinfo {author} {\bibfnamefont {T.}~\bibnamefont {Ishikawa}}, \bibinfo
  {author} {\bibfnamefont {T.~J.}\ \bibnamefont {Pedley}},\ and\ \bibinfo
  {author} {\bibfnamefont {R.~E.}\ \bibnamefont {Goldstein}},\ }\bibfield
  {title} {\bibinfo {title} {Dancing {V}olvox: Hydrodynamic bound states of
  swimming algae},\ }\href@noop {} {\bibfield  {journal} {\bibinfo  {journal}
  {Phys. Rev. Lett.}\ }\textbf {\bibinfo {volume} {102}},\ \bibinfo {pages}
  {168101} (\bibinfo {year} {2009})}\BibitemShut {NoStop}%
\bibitem [{\citenamefont {Llopis}\ and\ \citenamefont
  {Pagonabarraga}(2010)}]{llopis2010hydrodynamic}%
  \BibitemOpen
  \bibfield  {author} {\bibinfo {author} {\bibfnamefont {I.}~\bibnamefont
  {Llopis}}\ and\ \bibinfo {author} {\bibfnamefont {I.}~\bibnamefont
  {Pagonabarraga}},\ }\bibfield  {title} {\bibinfo {title} {Hydrodynamic
  interactions in squirmer motion: Swimming with a neighbour and close to a
  wall},\ }\href@noop {} {\bibfield  {journal} {\bibinfo  {journal} {Journal of
  Non-Newtonian Fluid Mechanics}\ }\textbf {\bibinfo {volume} {165}},\ \bibinfo
  {pages} {946} (\bibinfo {year} {2010})}\BibitemShut {NoStop}%
\bibitem [{\citenamefont {Ishimoto}\ and\ \citenamefont
  {Gaffney}(2013)}]{ishimoto2013squirmer}%
  \BibitemOpen
  \bibfield  {author} {\bibinfo {author} {\bibfnamefont {K.}~\bibnamefont
  {Ishimoto}}\ and\ \bibinfo {author} {\bibfnamefont {E.~A.}\ \bibnamefont
  {Gaffney}},\ }\bibfield  {title} {\bibinfo {title} {Squirmer dynamics near a
  boundary},\ }\href@noop {} {\bibfield  {journal} {\bibinfo  {journal}
  {Physical Review E}\ }\textbf {\bibinfo {volume} {88}},\ \bibinfo {pages}
  {062702} (\bibinfo {year} {2013})}\BibitemShut {NoStop}%
\bibitem [{\citenamefont {Li}\ and\ \citenamefont
  {Ardekani}(2014)}]{li2014hydrodynamic}%
  \BibitemOpen
  \bibfield  {author} {\bibinfo {author} {\bibfnamefont {G.-J.}\ \bibnamefont
  {Li}}\ and\ \bibinfo {author} {\bibfnamefont {A.~M.}\ \bibnamefont
  {Ardekani}},\ }\bibfield  {title} {\bibinfo {title} {Hydrodynamic interaction
  of microswimmers near a wall},\ }\href@noop {} {\bibfield  {journal}
  {\bibinfo  {journal} {Physical Review E}\ }\textbf {\bibinfo {volume} {90}},\
  \bibinfo {pages} {013010} (\bibinfo {year} {2014})}\BibitemShut {NoStop}%
\bibitem [{\citenamefont {Darveniza}\ \emph {et~al.}(2022)\citenamefont
  {Darveniza}, \citenamefont {Ishikawa}, \citenamefont {Pedley},\ and\
  \citenamefont {Brumley}}]{darveniza22}%
  \BibitemOpen
  \bibfield  {author} {\bibinfo {author} {\bibfnamefont {C.}~\bibnamefont
  {Darveniza}}, \bibinfo {author} {\bibfnamefont {T.}~\bibnamefont {Ishikawa}},
  \bibinfo {author} {\bibfnamefont {T.~J.}\ \bibnamefont {Pedley}},\ and\
  \bibinfo {author} {\bibfnamefont {D.~R.}\ \bibnamefont {Brumley}},\
  }\bibfield  {title} {\bibinfo {title} {Pairwise scattering and bound states
  of spherical microorganisms},\ }\href@noop {} {\bibfield  {journal} {\bibinfo
   {journal} {Phys. Rev. Fluids}\ }\textbf {\bibinfo {volume} {7}},\ \bibinfo
  {pages} {013104} (\bibinfo {year} {2022})}\BibitemShut {NoStop}%
\bibitem [{\citenamefont {Magar}\ \emph {et~al.}(2003)\citenamefont {Magar},
  \citenamefont {Goto},\ and\ \citenamefont {Pedley}}]{magar03}%
  \BibitemOpen
  \bibfield  {author} {\bibinfo {author} {\bibfnamefont {V.}~\bibnamefont
  {Magar}}, \bibinfo {author} {\bibfnamefont {T.}~\bibnamefont {Goto}},\ and\
  \bibinfo {author} {\bibfnamefont {T.~J.}\ \bibnamefont {Pedley}},\ }\bibfield
   {title} {\bibinfo {title} {Nutrient uptake by a self-propelled steady
  squirmer},\ }\href@noop {} {\bibfield  {journal} {\bibinfo  {journal} {The
  Quarterly Journal of Mechanics and Applied Mathematics}\ }\textbf {\bibinfo
  {volume} {56}},\ \bibinfo {pages} {65} (\bibinfo {year} {2003})}\BibitemShut
  {NoStop}%
\bibitem [{\citenamefont {Michelin}\ and\ \citenamefont
  {Lauga}(2011)}]{michelin2011}%
  \BibitemOpen
  \bibfield  {author} {\bibinfo {author} {\bibfnamefont {S.}~\bibnamefont
  {Michelin}}\ and\ \bibinfo {author} {\bibfnamefont {E.}~\bibnamefont
  {Lauga}},\ }\bibfield  {title} {\bibinfo {title} {Optimal feeding is optimal
  swimming for all p{\'e}clet numbers},\ }\href@noop {} {\bibfield  {journal}
  {\bibinfo  {journal} {Physics of Fluids}\ }\textbf {\bibinfo {volume} {23}},\
  \bibinfo {pages} {101901} (\bibinfo {year} {2011})}\BibitemShut {NoStop}%
\bibitem [{\citenamefont {Wang}\ \emph {et~al.}(2006)\citenamefont {Wang},
  \citenamefont {Hernandez}, \citenamefont {Bartlett}, \citenamefont {Bingham},
  \citenamefont {Kline}, \citenamefont {Sen},\ and\ \citenamefont
  {Mallouk}}]{wang2006bipolar}%
  \BibitemOpen
  \bibfield  {author} {\bibinfo {author} {\bibfnamefont {Y.}~\bibnamefont
  {Wang}}, \bibinfo {author} {\bibfnamefont {R.~M.}\ \bibnamefont {Hernandez}},
  \bibinfo {author} {\bibfnamefont {D.~J.}\ \bibnamefont {Bartlett}}, \bibinfo
  {author} {\bibfnamefont {J.~M.}\ \bibnamefont {Bingham}}, \bibinfo {author}
  {\bibfnamefont {T.~R.}\ \bibnamefont {Kline}}, \bibinfo {author}
  {\bibfnamefont {A.}~\bibnamefont {Sen}},\ and\ \bibinfo {author}
  {\bibfnamefont {T.~E.}\ \bibnamefont {Mallouk}},\ }\bibfield  {title}
  {\bibinfo {title} {Bipolar electrochemical mechanism for the propulsion of
  catalytic nanomotors in hydrogen peroxide solutions},\ }\href@noop {}
  {\bibfield  {journal} {\bibinfo  {journal} {Langmuir}\ }\textbf {\bibinfo
  {volume} {22}},\ \bibinfo {pages} {10451} (\bibinfo {year}
  {2006})}\BibitemShut {NoStop}%
\bibitem [{\citenamefont {Howse}\ \emph {et~al.}(2007)\citenamefont {Howse},
  \citenamefont {Jones}, \citenamefont {Ryan}, \citenamefont {Gough},
  \citenamefont {Vafabakhsh},\ and\ \citenamefont
  {Golestanian}}]{howse2007self}%
  \BibitemOpen
  \bibfield  {author} {\bibinfo {author} {\bibfnamefont {J.~R.}\ \bibnamefont
  {Howse}}, \bibinfo {author} {\bibfnamefont {R.~A.}\ \bibnamefont {Jones}},
  \bibinfo {author} {\bibfnamefont {A.~J.}\ \bibnamefont {Ryan}}, \bibinfo
  {author} {\bibfnamefont {T.}~\bibnamefont {Gough}}, \bibinfo {author}
  {\bibfnamefont {R.}~\bibnamefont {Vafabakhsh}},\ and\ \bibinfo {author}
  {\bibfnamefont {R.}~\bibnamefont {Golestanian}},\ }\bibfield  {title}
  {\bibinfo {title} {Self-motile colloidal particles: from directed propulsion
  to random walk},\ }\href@noop {} {\bibfield  {journal} {\bibinfo  {journal}
  {Physical review letters}\ }\textbf {\bibinfo {volume} {99}},\ \bibinfo
  {pages} {048102} (\bibinfo {year} {2007})}\BibitemShut {NoStop}%
\bibitem [{\citenamefont {Jiang}\ \emph {et~al.}(2010)\citenamefont {Jiang},
  \citenamefont {Yoshinaga},\ and\ \citenamefont {Sano}}]{jiang2010active}%
  \BibitemOpen
  \bibfield  {author} {\bibinfo {author} {\bibfnamefont {H.-R.}\ \bibnamefont
  {Jiang}}, \bibinfo {author} {\bibfnamefont {N.}~\bibnamefont {Yoshinaga}},\
  and\ \bibinfo {author} {\bibfnamefont {M.}~\bibnamefont {Sano}},\ }\bibfield
  {title} {\bibinfo {title} {Active motion of a {J}anus particle by
  self-thermophoresis in a defocused laser beam},\ }\href@noop {} {\bibfield
  {journal} {\bibinfo  {journal} {Physical review letters}\ }\textbf {\bibinfo
  {volume} {105}},\ \bibinfo {pages} {268302} (\bibinfo {year}
  {2010})}\BibitemShut {NoStop}%
\bibitem [{\citenamefont {Bregulla}\ and\ \citenamefont
  {Cichos}(2019)}]{bregulla2019flow}%
  \BibitemOpen
  \bibfield  {author} {\bibinfo {author} {\bibfnamefont {A.~P.}\ \bibnamefont
  {Bregulla}}\ and\ \bibinfo {author} {\bibfnamefont {F.}~\bibnamefont
  {Cichos}},\ }\bibfield  {title} {\bibinfo {title} {Flow fields around pinned
  self-thermophoretic microswimmers under confinement},\ }\href@noop {}
  {\bibfield  {journal} {\bibinfo  {journal} {The Journal of chemical physics}\
  }\textbf {\bibinfo {volume} {151}},\ \bibinfo {pages} {044706} (\bibinfo
  {year} {2019})}\BibitemShut {NoStop}%
\bibitem [{\citenamefont {Popescu}\ \emph
  {et~al.}(2018{\natexlab{a}})\citenamefont {Popescu}, \citenamefont {Uspal},
  \citenamefont {Eskandari}, \citenamefont {Tasinkevych},\ and\ \citenamefont
  {Dietrich}}]{Popescu18}%
  \BibitemOpen
  \bibfield  {author} {\bibinfo {author} {\bibfnamefont {M.}~\bibnamefont
  {Popescu}}, \bibinfo {author} {\bibfnamefont {W.}~\bibnamefont {Uspal}},
  \bibinfo {author} {\bibfnamefont {Z.}~\bibnamefont {Eskandari}}, \bibinfo
  {author} {\bibfnamefont {M.}~\bibnamefont {Tasinkevych}},\ and\ \bibinfo
  {author} {\bibfnamefont {S.}~\bibnamefont {Dietrich}},\ }\bibfield  {title}
  {\bibinfo {title} {Effective squirmer models for self-phoretic chemically
  active spherical colloids},\ }\href@noop {} {\bibfield  {journal} {\bibinfo
  {journal} {Eur. Phys. J. E}\ }\textbf {\bibinfo {volume} {41}},\ \bibinfo
  {pages} {145} (\bibinfo {year} {2018}{\natexlab{a}})}\BibitemShut {NoStop}%
\bibitem [{\citenamefont {Lauga}\ and\ \citenamefont
  {Michelin}(2016)}]{Lauga16}%
  \BibitemOpen
  \bibfield  {author} {\bibinfo {author} {\bibfnamefont {E.}~\bibnamefont
  {Lauga}}\ and\ \bibinfo {author} {\bibfnamefont {S.}~\bibnamefont
  {Michelin}},\ }\bibfield  {title} {\bibinfo {title} {Stresslets induced by
  active swimmers},\ }\href@noop {} {\bibfield  {journal} {\bibinfo  {journal}
  {Phys. Rev. Lett.}\ }\textbf {\bibinfo {volume} {117}},\ \bibinfo {pages}
  {148001} (\bibinfo {year} {2016})}\BibitemShut {NoStop}%
\bibitem [{\citenamefont {Stone}\ and\ \citenamefont {Samuel}(1996)}]{Stone96}%
  \BibitemOpen
  \bibfield  {author} {\bibinfo {author} {\bibfnamefont {H.~A.}\ \bibnamefont
  {Stone}}\ and\ \bibinfo {author} {\bibfnamefont {A.~D.~T.}\ \bibnamefont
  {Samuel}},\ }\bibfield  {title} {\bibinfo {title} {Propulsion of
  microorganisms by surface distortions},\ }\href@noop {} {\bibfield  {journal}
  {\bibinfo  {journal} {Phys. Rev. Lett.}\ }\textbf {\bibinfo {volume} {77}},\
  \bibinfo {pages} {4102} (\bibinfo {year} {1996})}\BibitemShut {NoStop}%
\bibitem [{\citenamefont {Katuri}\ \emph {et~al.}(2022)\citenamefont {Katuri},
  \citenamefont {Poehnl}, \citenamefont {Sokolov}, \citenamefont {Uspal},\ and\
  \citenamefont {Snezhko}}]{katuri22}%
  \BibitemOpen
  \bibfield  {author} {\bibinfo {author} {\bibfnamefont {J.}~\bibnamefont
  {Katuri}}, \bibinfo {author} {\bibfnamefont {R.}~\bibnamefont {Poehnl}},
  \bibinfo {author} {\bibfnamefont {A.}~\bibnamefont {Sokolov}}, \bibinfo
  {author} {\bibfnamefont {W.}~\bibnamefont {Uspal}},\ and\ \bibinfo {author}
  {\bibfnamefont {A.}~\bibnamefont {Snezhko}},\ }\bibfield  {title} {\bibinfo
  {title} {Arrested-motility states in populations of shape-anisotropic active
  {J}anus particles},\ }\href {https://doi.org/10.1126/sciadv.abo3604}
  {\bibfield  {journal} {\bibinfo  {journal} {Science Advances}\ }\textbf
  {\bibinfo {volume} {8}},\ \bibinfo {pages} {eabo3604} (\bibinfo {year}
  {2022})}\BibitemShut {NoStop}%
\bibitem [{\citenamefont {Squires}\ and\ \citenamefont
  {Bazant}(2006)}]{squires06}%
  \BibitemOpen
  \bibfield  {author} {\bibinfo {author} {\bibfnamefont {T.~M.}\ \bibnamefont
  {Squires}}\ and\ \bibinfo {author} {\bibfnamefont {M.~Z.}\ \bibnamefont
  {Bazant}},\ }\bibfield  {title} {\bibinfo {title} {Breaking symmetries in
  induced-charge electro-osmosis and electrophoresis},\ }\href@noop {}
  {\bibfield  {journal} {\bibinfo  {journal} {J. Fluid Mech.}\ }\textbf
  {\bibinfo {volume} {560}},\ \bibinfo {pages} {65} (\bibinfo {year}
  {2006})}\BibitemShut {NoStop}%
\bibitem [{\citenamefont {Kilic}\ and\ \citenamefont {Bazant}(2011)}]{kilic11}%
  \BibitemOpen
  \bibfield  {author} {\bibinfo {author} {\bibfnamefont {M.~S.}\ \bibnamefont
  {Kilic}}\ and\ \bibinfo {author} {\bibfnamefont {M.~Z.}\ \bibnamefont
  {Bazant}},\ }\bibfield  {title} {\bibinfo {title} {Induced-charge
  electrophoresis near a wall},\ }\href@noop {} {\bibfield  {journal} {\bibinfo
   {journal} {Electrophoresis}\ }\textbf {\bibinfo {volume} {32}},\ \bibinfo
  {pages} {614–628} (\bibinfo {year} {2011})}\BibitemShut {NoStop}%
\bibitem [{\citenamefont {Brooks}\ \emph {et~al.}(2018)\citenamefont {Brooks},
  \citenamefont {Sabrina},\ and\ \citenamefont {Bishop}}]{brooks2018shape}%
  \BibitemOpen
  \bibfield  {author} {\bibinfo {author} {\bibfnamefont {A.~M.}\ \bibnamefont
  {Brooks}}, \bibinfo {author} {\bibfnamefont {S.}~\bibnamefont {Sabrina}},\
  and\ \bibinfo {author} {\bibfnamefont {K.~J.}\ \bibnamefont {Bishop}},\
  }\bibfield  {title} {\bibinfo {title} {Shape-directed dynamics of active
  colloids powered by induced-charge electrophoresis},\ }\href@noop {}
  {\bibfield  {journal} {\bibinfo  {journal} {Proceedings of the national
  academy of sciences}\ }\textbf {\bibinfo {volume} {115}},\ \bibinfo {pages}
  {E1090} (\bibinfo {year} {2018})}\BibitemShut {NoStop}%
\bibitem [{\citenamefont {Sharan}\ \emph {et~al.}(2021)\citenamefont {Sharan},
  \citenamefont {Maslen}, \citenamefont {Altunkeyik}, \citenamefont {Rehor},
  \citenamefont {Simmchen},\ and\ \citenamefont
  {Montenegro-Johnson}}]{sharan2021fundamental}%
  \BibitemOpen
  \bibfield  {author} {\bibinfo {author} {\bibfnamefont {P.}~\bibnamefont
  {Sharan}}, \bibinfo {author} {\bibfnamefont {C.}~\bibnamefont {Maslen}},
  \bibinfo {author} {\bibfnamefont {B.}~\bibnamefont {Altunkeyik}}, \bibinfo
  {author} {\bibfnamefont {I.}~\bibnamefont {Rehor}}, \bibinfo {author}
  {\bibfnamefont {J.}~\bibnamefont {Simmchen}},\ and\ \bibinfo {author}
  {\bibfnamefont {T.~D.}\ \bibnamefont {Montenegro-Johnson}},\ }\bibfield
  {title} {\bibinfo {title} {Fundamental modes of swimming correspond to
  fundamental modes of shape: Engineering i-, u-, and s-shaped swimmers},\
  }\href@noop {} {\bibfield  {journal} {\bibinfo  {journal} {Advanced
  Intelligent Systems}\ }\textbf {\bibinfo {volume} {3}},\ \bibinfo {pages}
  {2100068} (\bibinfo {year} {2021})}\BibitemShut {NoStop}%
\bibitem [{\citenamefont {Diwakar}\ \emph {et~al.}(2022)\citenamefont
  {Diwakar}, \citenamefont {Kunti}, \citenamefont {Miloh}, \citenamefont
  {Yossifon},\ and\ \citenamefont {Velev}}]{diwakar2022}%
  \BibitemOpen
  \bibfield  {author} {\bibinfo {author} {\bibfnamefont {N.~M.}\ \bibnamefont
  {Diwakar}}, \bibinfo {author} {\bibfnamefont {G.}~\bibnamefont {Kunti}},
  \bibinfo {author} {\bibfnamefont {T.}~\bibnamefont {Miloh}}, \bibinfo
  {author} {\bibfnamefont {G.}~\bibnamefont {Yossifon}},\ and\ \bibinfo
  {author} {\bibfnamefont {O.~D.}\ \bibnamefont {Velev}},\ }\bibfield  {title}
  {\bibinfo {title} {Ac electrohydrodynamic propulsion and rotation of active
  particles of engineered shape and asymmetry},\ }\href@noop {} {\bibfield
  {journal} {\bibinfo  {journal} {Current Opinion in Colloid \& Interface
  Science}\ ,\ \bibinfo {pages} {101586}} (\bibinfo {year} {2022})}\BibitemShut
  {NoStop}%
\bibitem [{\citenamefont {Ganguly}\ and\ \citenamefont
  {Gupta}(2023)}]{ganguly2023going}%
  \BibitemOpen
  \bibfield  {author} {\bibinfo {author} {\bibfnamefont {A.}~\bibnamefont
  {Ganguly}}\ and\ \bibinfo {author} {\bibfnamefont {A.}~\bibnamefont
  {Gupta}},\ }\bibfield  {title} {\bibinfo {title} {Going in circles: Slender
  body analysis of a self-propelling bent rod},\ }\href@noop {} {\bibfield
  {journal} {\bibinfo  {journal} {Physical Review Fluids}\ }\textbf {\bibinfo
  {volume} {8}},\ \bibinfo {pages} {014103} (\bibinfo {year}
  {2023})}\BibitemShut {NoStop}%
\bibitem [{\citenamefont {Theers}\ \emph {et~al.}(2016)\citenamefont {Theers},
  \citenamefont {Westphal}, \citenamefont {Gompper},\ and\ \citenamefont
  {Winkler}}]{theers16}%
  \BibitemOpen
  \bibfield  {author} {\bibinfo {author} {\bibfnamefont {M.}~\bibnamefont
  {Theers}}, \bibinfo {author} {\bibfnamefont {E.}~\bibnamefont {Westphal}},
  \bibinfo {author} {\bibfnamefont {G.}~\bibnamefont {Gompper}},\ and\ \bibinfo
  {author} {\bibfnamefont {R.~G.}\ \bibnamefont {Winkler}},\ }\bibfield
  {title} {\bibinfo {title} {Modeling a spheroidal microswimmer and cooperative
  swimming in a narrow slit},\ }\href@noop {} {\bibfield  {journal} {\bibinfo
  {journal} {Soft Matter}\ }\textbf {\bibinfo {volume} {12}},\ \bibinfo {pages}
  {7372} (\bibinfo {year} {2016})}\BibitemShut {NoStop}%
\bibitem [{\citenamefont {Michelin}\ and\ \citenamefont
  {Lauga}(2017)}]{michelin2017geometric}%
  \BibitemOpen
  \bibfield  {author} {\bibinfo {author} {\bibfnamefont {S.}~\bibnamefont
  {Michelin}}\ and\ \bibinfo {author} {\bibfnamefont {E.}~\bibnamefont
  {Lauga}},\ }\bibfield  {title} {\bibinfo {title} {Geometric tuning of
  self-propulsion for {J}anus catalytic particles},\ }\href@noop {} {\bibfield
  {journal} {\bibinfo  {journal} {Scientific reports}\ }\textbf {\bibinfo
  {volume} {7}},\ \bibinfo {pages} {1} (\bibinfo {year} {2017})}\BibitemShut
  {NoStop}%
\bibitem [{\citenamefont {Yariv}(2019)}]{yariv2019self}%
  \BibitemOpen
  \bibfield  {author} {\bibinfo {author} {\bibfnamefont {E.}~\bibnamefont
  {Yariv}},\ }\bibfield  {title} {\bibinfo {title} {Self-diffusiophoresis of
  slender catalytic colloids},\ }\href@noop {} {\bibfield  {journal} {\bibinfo
  {journal} {Langmuir}\ }\textbf {\bibinfo {volume} {36}},\ \bibinfo {pages}
  {6903} (\bibinfo {year} {2019})}\BibitemShut {NoStop}%
\bibitem [{\citenamefont {Daddi-Moussa-Ider}\ \emph {et~al.}(2021)\citenamefont
  {Daddi-Moussa-Ider}, \citenamefont {Nasouri}, \citenamefont {Vilfan},\ and\
  \citenamefont {Golestanian}}]{daddi2021optimal}%
  \BibitemOpen
  \bibfield  {author} {\bibinfo {author} {\bibfnamefont {A.}~\bibnamefont
  {Daddi-Moussa-Ider}}, \bibinfo {author} {\bibfnamefont {B.}~\bibnamefont
  {Nasouri}}, \bibinfo {author} {\bibfnamefont {A.}~\bibnamefont {Vilfan}},\
  and\ \bibinfo {author} {\bibfnamefont {R.}~\bibnamefont {Golestanian}},\
  }\bibfield  {title} {\bibinfo {title} {Optimal swimmers can be pullers,
  pushers or neutral depending on the shape},\ }\href@noop {} {\bibfield
  {journal} {\bibinfo  {journal} {Journal of Fluid Mechanics}\ }\textbf
  {\bibinfo {volume} {922}},\ \bibinfo {pages} {R5} (\bibinfo {year}
  {2021})}\BibitemShut {NoStop}%
\bibitem [{\citenamefont {Poehnl}\ and\ \citenamefont
  {Uspal}(2021)}]{poehnl21}%
  \BibitemOpen
  \bibfield  {author} {\bibinfo {author} {\bibfnamefont {R.}~\bibnamefont
  {Poehnl}}\ and\ \bibinfo {author} {\bibfnamefont {W.}~\bibnamefont {Uspal}},\
  }\bibfield  {title} {\bibinfo {title} {Phoretic self-propulsion of helical
  active particles},\ }\href@noop {} {\bibfield  {journal} {\bibinfo  {journal}
  {Journal of Fluid Mechanics}\ }\textbf {\bibinfo {volume} {927}},\ \bibinfo
  {pages} {A46} (\bibinfo {year} {2021})}\BibitemShut {NoStop}%
\bibitem [{\citenamefont {Zantop}\ and\ \citenamefont
  {Stark}(2022)}]{zantop2022emergent}%
  \BibitemOpen
  \bibfield  {author} {\bibinfo {author} {\bibfnamefont {A.~W.}\ \bibnamefont
  {Zantop}}\ and\ \bibinfo {author} {\bibfnamefont {H.}~\bibnamefont {Stark}},\
  }\bibfield  {title} {\bibinfo {title} {Emergent collective dynamics of pusher
  and puller squirmer rods: swarming, clustering, and turbulence},\ }\href@noop
  {} {\bibfield  {journal} {\bibinfo  {journal} {Soft Matter}\ }\textbf
  {\bibinfo {volume} {18}},\ \bibinfo {pages} {6179} (\bibinfo {year}
  {2022})}\BibitemShut {NoStop}%
\bibitem [{\citenamefont {B{\"a}r}\ \emph {et~al.}(2020)\citenamefont
  {B{\"a}r}, \citenamefont {Gro{\ss}mann}, \citenamefont {Heidenreich},\ and\
  \citenamefont {Peruani}}]{bar2020self}%
  \BibitemOpen
  \bibfield  {author} {\bibinfo {author} {\bibfnamefont {M.}~\bibnamefont
  {B{\"a}r}}, \bibinfo {author} {\bibfnamefont {R.}~\bibnamefont
  {Gro{\ss}mann}}, \bibinfo {author} {\bibfnamefont {S.}~\bibnamefont
  {Heidenreich}},\ and\ \bibinfo {author} {\bibfnamefont {F.}~\bibnamefont
  {Peruani}},\ }\bibfield  {title} {\bibinfo {title} {Self-propelled rods:
  Insights and perspectives for active matter},\ }\href@noop {} {\bibfield
  {journal} {\bibinfo  {journal} {Annual Review of Condensed Matter Physics}\
  }\textbf {\bibinfo {volume} {11}},\ \bibinfo {pages} {441} (\bibinfo {year}
  {2020})}\BibitemShut {NoStop}%
\bibitem [{\citenamefont {Felderhof}(2016)}]{felderhof2016stokesian}%
  \BibitemOpen
  \bibfield  {author} {\bibinfo {author} {\bibfnamefont {B.}~\bibnamefont
  {Felderhof}},\ }\bibfield  {title} {\bibinfo {title} {Stokesian swimming of a
  prolate spheroid at low {R}eynolds number},\ }\href@noop {} {\bibfield
  {journal} {\bibinfo  {journal} {European Journal of Mechanics-B/Fluids}\
  }\textbf {\bibinfo {volume} {60}},\ \bibinfo {pages} {230} (\bibinfo {year}
  {2016})}\BibitemShut {NoStop}%
\bibitem [{\citenamefont {Leshansky}\ \emph {et~al.}(2007)\citenamefont
  {Leshansky}, \citenamefont {Kenneth}, \citenamefont {Gat},\ and\
  \citenamefont {Avron}}]{leshansky2007frictionless}%
  \BibitemOpen
  \bibfield  {author} {\bibinfo {author} {\bibfnamefont {A.~M.}\ \bibnamefont
  {Leshansky}}, \bibinfo {author} {\bibfnamefont {O.}~\bibnamefont {Kenneth}},
  \bibinfo {author} {\bibfnamefont {O.}~\bibnamefont {Gat}},\ and\ \bibinfo
  {author} {\bibfnamefont {J.~E.}\ \bibnamefont {Avron}},\ }\bibfield  {title}
  {\bibinfo {title} {A frictionless microswimmer},\ }\href@noop {} {\bibfield
  {journal} {\bibinfo  {journal} {New Journal of Physics}\ }\textbf {\bibinfo
  {volume} {9}},\ \bibinfo {pages} {145} (\bibinfo {year} {2007})}\BibitemShut
  {NoStop}%
\bibitem [{\citenamefont {Ishikawa}\ and\ \citenamefont
  {Hota}(2006)}]{ishikawa2006interaction}%
  \BibitemOpen
  \bibfield  {author} {\bibinfo {author} {\bibfnamefont {T.}~\bibnamefont
  {Ishikawa}}\ and\ \bibinfo {author} {\bibfnamefont {M.}~\bibnamefont
  {Hota}},\ }\bibfield  {title} {\bibinfo {title} {Interaction of two swimming
  paramecia},\ }\href@noop {} {\bibfield  {journal} {\bibinfo  {journal}
  {Journal of Experimental Biology}\ }\textbf {\bibinfo {volume} {209}},\
  \bibinfo {pages} {4452} (\bibinfo {year} {2006})}\BibitemShut {NoStop}%
\bibitem [{\citenamefont {P\"ohnl}\ \emph {et~al.}(2020)\citenamefont
  {P\"ohnl}, \citenamefont {Popescu},\ and\ \citenamefont {Uspal}}]{poehnl20}%
  \BibitemOpen
  \bibfield  {author} {\bibinfo {author} {\bibfnamefont {R.}~\bibnamefont
  {P\"ohnl}}, \bibinfo {author} {\bibfnamefont {M.~N.}\ \bibnamefont
  {Popescu}},\ and\ \bibinfo {author} {\bibfnamefont {W.~E.}\ \bibnamefont
  {Uspal}},\ }\bibfield  {title} {\bibinfo {title} {Axisymmetric spheroidal
  squirmers and self-diffusiophoretic particles},\ }\href@noop {} {\bibfield
  {journal} {\bibinfo  {journal} {J. Phys.: Condens. Matter}\ }\textbf
  {\bibinfo {volume} {32}},\ \bibinfo {pages} {164001} (\bibinfo {year}
  {2020})}\BibitemShut {NoStop}%
\bibitem [{\citenamefont {Archer}\ \emph {et~al.}(2015)\citenamefont {Archer},
  \citenamefont {Campbell},\ and\ \citenamefont {Ebbens}}]{archer2015}%
  \BibitemOpen
  \bibfield  {author} {\bibinfo {author} {\bibfnamefont {R.}~\bibnamefont
  {Archer}}, \bibinfo {author} {\bibfnamefont {A.}~\bibnamefont {Campbell}},\
  and\ \bibinfo {author} {\bibfnamefont {S.}~\bibnamefont {Ebbens}},\
  }\bibfield  {title} {\bibinfo {title} {Glancing angle metal evaporation
  synthesis of catalytic swimming {J}anus colloids with well defined angular
  velocity},\ }\href@noop {} {\bibfield  {journal} {\bibinfo  {journal} {Soft
  Matter}\ }\textbf {\bibinfo {volume} {11}},\ \bibinfo {pages} {6872}
  (\bibinfo {year} {2015})}\BibitemShut {NoStop}%
\bibitem [{\citenamefont {Lisicki}\ \emph {et~al.}(2018)\citenamefont
  {Lisicki}, \citenamefont {Reigh},\ and\ \citenamefont {Lauga}}]{Lisicki18}%
  \BibitemOpen
  \bibfield  {author} {\bibinfo {author} {\bibfnamefont {M.}~\bibnamefont
  {Lisicki}}, \bibinfo {author} {\bibfnamefont {S.~Y.}\ \bibnamefont {Reigh}},\
  and\ \bibinfo {author} {\bibfnamefont {E.}~\bibnamefont {Lauga}},\ }\bibfield
   {title} {\bibinfo {title} {Autophoretic motion in three dimensions},\
  }\href@noop {} {\bibfield  {journal} {\bibinfo  {journal} {Soft Matter}\
  }\textbf {\bibinfo {volume} {14}},\ \bibinfo {pages} {3304} (\bibinfo {year}
  {2018})}\BibitemShut {NoStop}%
\bibitem [{\citenamefont {Uspal}\ \emph {et~al.}(2015)\citenamefont {Uspal},
  \citenamefont {Popescu}, \citenamefont {Dietrich},\ and\ \citenamefont
  {Tasinkevych}}]{uspal15}%
  \BibitemOpen
  \bibfield  {author} {\bibinfo {author} {\bibfnamefont {W.~E.}\ \bibnamefont
  {Uspal}}, \bibinfo {author} {\bibfnamefont {M.~N.}\ \bibnamefont {Popescu}},
  \bibinfo {author} {\bibfnamefont {S.}~\bibnamefont {Dietrich}},\ and\
  \bibinfo {author} {\bibfnamefont {M.}~\bibnamefont {Tasinkevych}},\
  }\bibfield  {title} {\bibinfo {title} {Self-propulsion of a catalytically
  active particle near a planar wall: from reflection to sliding and
  hovering},\ }\href@noop {} {\bibfield  {journal} {\bibinfo  {journal} {Soft
  Matter}\ }\textbf {\bibinfo {volume} {11}},\ \bibinfo {pages} {434} (\bibinfo
  {year} {2015})}\BibitemShut {NoStop}%
\bibitem [{\citenamefont {Popescu}\ \emph
  {et~al.}(2018{\natexlab{b}})\citenamefont {Popescu}, \citenamefont {Uspal},
  \citenamefont {Bechinger},\ and\ \citenamefont
  {Fischer}}]{popescu2018chemotaxis}%
  \BibitemOpen
  \bibfield  {author} {\bibinfo {author} {\bibfnamefont {M.~N.}\ \bibnamefont
  {Popescu}}, \bibinfo {author} {\bibfnamefont {W.~E.}\ \bibnamefont {Uspal}},
  \bibinfo {author} {\bibfnamefont {C.}~\bibnamefont {Bechinger}},\ and\
  \bibinfo {author} {\bibfnamefont {P.}~\bibnamefont {Fischer}},\ }\bibfield
  {title} {\bibinfo {title} {Chemotaxis of active {J}anus nanoparticles},\
  }\href@noop {} {\bibfield  {journal} {\bibinfo  {journal} {Nano letters}\
  }\textbf {\bibinfo {volume} {18}},\ \bibinfo {pages} {5345} (\bibinfo {year}
  {2018}{\natexlab{b}})}\BibitemShut {NoStop}%
\bibitem [{\citenamefont {Ghose}\ and\ \citenamefont
  {Adhikari}(2014)}]{Ghose14}%
  \BibitemOpen
  \bibfield  {author} {\bibinfo {author} {\bibfnamefont {S.}~\bibnamefont
  {Ghose}}\ and\ \bibinfo {author} {\bibfnamefont {R.}~\bibnamefont
  {Adhikari}},\ }\bibfield  {title} {\bibinfo {title} {Irreducible
  representations of oscillatory and swirling flows in active soft matter},\
  }\href@noop {} {\bibfield  {journal} {\bibinfo  {journal} {Phys. Rev. Lett.}\
  }\textbf {\bibinfo {volume} {112}},\ \bibinfo {pages} {118102} (\bibinfo
  {year} {2014})}\BibitemShut {NoStop}%
\bibitem [{\citenamefont {Pak}\ and\ \citenamefont {Lauga}(2014)}]{Pak14}%
  \BibitemOpen
  \bibfield  {author} {\bibinfo {author} {\bibfnamefont {O.}~\bibnamefont
  {Pak}}\ and\ \bibinfo {author} {\bibfnamefont {E.}~\bibnamefont {Lauga}},\
  }\bibfield  {title} {\bibinfo {title} {Generalized squirming motion of a
  sphere},\ }\href@noop {} {\bibfield  {journal} {\bibinfo  {journal} {J. Eng.
  Math.}\ }\textbf {\bibinfo {volume} {88}},\ \bibinfo {pages} {1} (\bibinfo
  {year} {2014})}\BibitemShut {NoStop}%
\bibitem [{\citenamefont {Felderhof}\ and\ \citenamefont
  {Jones}(2016)}]{felderhof2016}%
  \BibitemOpen
  \bibfield  {author} {\bibinfo {author} {\bibfnamefont {B.}~\bibnamefont
  {Felderhof}}\ and\ \bibinfo {author} {\bibfnamefont {R.}~\bibnamefont
  {Jones}},\ }\bibfield  {title} {\bibinfo {title} {Stokesian swimming of a
  sphere at low {R}eynolds number by helical surface distortion},\ }\href@noop
  {} {\bibfield  {journal} {\bibinfo  {journal} {Physics of Fluids}\ }\textbf
  {\bibinfo {volume} {28}},\ \bibinfo {pages} {073601} (\bibinfo {year}
  {2016})}\BibitemShut {NoStop}%
\bibitem [{\citenamefont {Pedley}\ \emph {et~al.}(2016)\citenamefont {Pedley},
  \citenamefont {Brumley},\ and\ \citenamefont {Goldstein}}]{pedley16}%
  \BibitemOpen
  \bibfield  {author} {\bibinfo {author} {\bibfnamefont {T.~J.}\ \bibnamefont
  {Pedley}}, \bibinfo {author} {\bibfnamefont {D.~R.}\ \bibnamefont
  {Brumley}},\ and\ \bibinfo {author} {\bibfnamefont {R.~E.}\ \bibnamefont
  {Goldstein}},\ }\bibfield  {title} {\bibinfo {title} {Squirmers with swirl: a
  model for volvox swimming},\ }\href@noop {} {\bibfield  {journal} {\bibinfo
  {journal} {Journal of Fluid Mechanics}\ }\textbf {\bibinfo {volume} {798}},\
  \bibinfo {pages} {165–186} (\bibinfo {year} {2016})}\BibitemShut {NoStop}%
\bibitem [{\citenamefont {Burada}\ \emph {et~al.}(2022)\citenamefont {Burada},
  \citenamefont {Maity},\ and\ \citenamefont {J{\"u}licher}}]{burada2022}%
  \BibitemOpen
  \bibfield  {author} {\bibinfo {author} {\bibfnamefont {P.}~\bibnamefont
  {Burada}}, \bibinfo {author} {\bibfnamefont {R.}~\bibnamefont {Maity}},\ and\
  \bibinfo {author} {\bibfnamefont {F.}~\bibnamefont {J{\"u}licher}},\
  }\bibfield  {title} {\bibinfo {title} {Hydrodynamics of chiral squirmers},\
  }\href@noop {} {\bibfield  {journal} {\bibinfo  {journal} {Physical Review
  E}\ }\textbf {\bibinfo {volume} {105}},\ \bibinfo {pages} {024603} (\bibinfo
  {year} {2022})}\BibitemShut {NoStop}%
\bibitem [{\citenamefont {Saintillan}\ and\ \citenamefont
  {Shelley}(2007)}]{saintillan2007orientational}%
  \BibitemOpen
  \bibfield  {author} {\bibinfo {author} {\bibfnamefont {D.}~\bibnamefont
  {Saintillan}}\ and\ \bibinfo {author} {\bibfnamefont {M.~J.}\ \bibnamefont
  {Shelley}},\ }\bibfield  {title} {\bibinfo {title} {Orientational order and
  instabilities in suspensions of self-locomoting rods},\ }\href@noop {}
  {\bibfield  {journal} {\bibinfo  {journal} {Physical review letters}\
  }\textbf {\bibinfo {volume} {99}},\ \bibinfo {pages} {058102} (\bibinfo
  {year} {2007})}\BibitemShut {NoStop}%
\bibitem [{\citenamefont {Saintillan}\ and\ \citenamefont
  {Shelley}(2008)}]{saintillan2008}%
  \BibitemOpen
  \bibfield  {author} {\bibinfo {author} {\bibfnamefont {D.}~\bibnamefont
  {Saintillan}}\ and\ \bibinfo {author} {\bibfnamefont {M.~J.}\ \bibnamefont
  {Shelley}},\ }\bibfield  {title} {\bibinfo {title} {Instabilities, pattern
  formation, and mixing in active suspensions},\ }\href@noop {} {\bibfield
  {journal} {\bibinfo  {journal} {Physics of Fluids}\ }\textbf {\bibinfo
  {volume} {20}} (\bibinfo {year} {2008})}\BibitemShut {NoStop}%
\bibitem [{\citenamefont {Saintillan}\ and\ \citenamefont
  {Shelley}(2013)}]{Saintillan13}%
  \BibitemOpen
  \bibfield  {author} {\bibinfo {author} {\bibfnamefont {D.}~\bibnamefont
  {Saintillan}}\ and\ \bibinfo {author} {\bibfnamefont {M.}~\bibnamefont
  {Shelley}},\ }\bibfield  {title} {\bibinfo {title} {Active suspensions and
  their nonlinear models},\ }\href@noop {} {\bibfield  {journal} {\bibinfo
  {journal} {Comptes Rendus Physique}\ }\textbf {\bibinfo {volume} {14}},\
  \bibinfo {pages} {497} (\bibinfo {year} {2013})}\BibitemShut {NoStop}%
\bibitem [{\citenamefont {Lushi}\ and\ \citenamefont
  {Peskin}(2013)}]{lushi2013modeling}%
  \BibitemOpen
  \bibfield  {author} {\bibinfo {author} {\bibfnamefont {E.}~\bibnamefont
  {Lushi}}\ and\ \bibinfo {author} {\bibfnamefont {C.~S.}\ \bibnamefont
  {Peskin}},\ }\bibfield  {title} {\bibinfo {title} {Modeling and simulation of
  active suspensions containing large numbers of interacting micro-swimmers},\
  }\href@noop {} {\bibfield  {journal} {\bibinfo  {journal} {Computers \&
  Structures}\ }\textbf {\bibinfo {volume} {122}},\ \bibinfo {pages} {239}
  (\bibinfo {year} {2013})}\BibitemShut {NoStop}%
\bibitem [{\citenamefont {Kim}\ and\ \citenamefont
  {Karrila}(2013)}]{kim2013microhydrodynamics}%
  \BibitemOpen
  \bibfield  {author} {\bibinfo {author} {\bibfnamefont {S.}~\bibnamefont
  {Kim}}\ and\ \bibinfo {author} {\bibfnamefont {S.~J.}\ \bibnamefont
  {Karrila}},\ }\href@noop {} {\emph {\bibinfo {title} {Microhydrodynamics:
  principles and selected applications}}}\ (\bibinfo  {publisher} {Courier
  Corporation},\ \bibinfo {year} {2013})\BibitemShut {NoStop}%
\bibitem [{\citenamefont {Saintillan}(2017)}]{Saintillan17}%
  \BibitemOpen
  \bibfield  {author} {\bibinfo {author} {\bibfnamefont {D.}~\bibnamefont
  {Saintillan}},\ }\bibfield  {title} {\bibinfo {title} {Rheology of active
  fluids},\ }\href@noop {} {\bibfield  {journal} {\bibinfo  {journal} {Ann.
  Rev. Fluid Mech.}\ }\textbf {\bibinfo {volume} {50}},\ \bibinfo {pages}
  {563–92} (\bibinfo {year} {2017})}\BibitemShut {NoStop}%
\bibitem [{Note1()}]{Note1}%
  \BibitemOpen
  \bibinfo {note} {See Supplemental Material at \protect \url
  {http://link.aps.org/supplemental/XXX} for technical discussion of the
  stresslet tensor, detailed mathematical derivations of results given in the
  main text, details concerning implementation of the spheroidal squirmer
  model, and numerical results obtained with the spheroidal squirmer model. The
  Supplemental Material also contains Refs. \protect \citenum
  {kilic11,poehnl20,katuri22,dassios1994generalized}.}\BibitemShut {Stop}%
\bibitem [{\citenamefont {Pozrikidis}(2002)}]{pozrikidis02}%
  \BibitemOpen
  \bibfield  {author} {\bibinfo {author} {\bibfnamefont {C.}~\bibnamefont
  {Pozrikidis}},\ }\href@noop {} {\emph {\bibinfo {title} {A Practical Guide to
  Boundary Element Methods with the Software Library BEMLIB}}}\ (\bibinfo
  {publisher} {CRC Press},\ \bibinfo {address} {Boca Raton},\ \bibinfo {year}
  {2002})\BibitemShut {NoStop}%
\bibitem [{\citenamefont {Jeffery}(1922)}]{jeffery1922motion}%
  \BibitemOpen
  \bibfield  {author} {\bibinfo {author} {\bibfnamefont {G.~B.}\ \bibnamefont
  {Jeffery}},\ }\bibfield  {title} {\bibinfo {title} {The motion of ellipsoidal
  particles immersed in a viscous fluid},\ }\href@noop {} {\bibfield  {journal}
  {\bibinfo  {journal} {Proceedings of the Royal Society of London. Series A,
  Containing papers of a mathematical and physical character}\ }\textbf
  {\bibinfo {volume} {102}},\ \bibinfo {pages} {161} (\bibinfo {year}
  {1922})}\BibitemShut {NoStop}%
\bibitem [{\citenamefont {Graham}(2018)}]{graham2018microhydrodynamics}%
  \BibitemOpen
  \bibfield  {author} {\bibinfo {author} {\bibfnamefont {M.~D.}\ \bibnamefont
  {Graham}},\ }\href@noop {} {\emph {\bibinfo {title} {Microhydrodynamics,
  Brownian motion, and complex fluids}}},\ Vol.~\bibinfo {volume} {58}\
  (\bibinfo  {publisher} {Cambridge University Press},\ \bibinfo {year}
  {2018})\BibitemShut {NoStop}%
\bibitem [{\citenamefont {Ouyang}\ \emph {et~al.}(2023)\citenamefont {Ouyang},
  \citenamefont {Lin}, \citenamefont {Lin}, \citenamefont {Yu},\ and\
  \citenamefont {Phan-Thien}}]{ouyang23}%
  \BibitemOpen
  \bibfield  {author} {\bibinfo {author} {\bibfnamefont {Z.}~\bibnamefont
  {Ouyang}}, \bibinfo {author} {\bibfnamefont {Z.}~\bibnamefont {Lin}},
  \bibinfo {author} {\bibfnamefont {J.}~\bibnamefont {Lin}}, \bibinfo {author}
  {\bibfnamefont {Z.}~\bibnamefont {Yu}},\ and\ \bibinfo {author}
  {\bibfnamefont {N.}~\bibnamefont {Phan-Thien}},\ }\bibfield  {title}
  {\bibinfo {title} {Cargo carrying with an inertial squirmer in a newtonian
  fluid},\ }\href@noop {} {\bibfield  {journal} {\bibinfo  {journal} {Journal
  of Fluid Mechanics}\ }\textbf {\bibinfo {volume} {959}},\ \bibinfo {pages}
  {A25} (\bibinfo {year} {2023})}\BibitemShut {NoStop}%
\bibitem [{\citenamefont {Claeys}\ and\ \citenamefont
  {Brady}(1993)}]{claeys1993suspensions}%
  \BibitemOpen
  \bibfield  {author} {\bibinfo {author} {\bibfnamefont {I.~L.}\ \bibnamefont
  {Claeys}}\ and\ \bibinfo {author} {\bibfnamefont {J.~F.}\ \bibnamefont
  {Brady}},\ }\bibfield  {title} {\bibinfo {title} {Suspensions of prolate
  spheroids in stokes flow. part 1. dynamics of a finite number of particles in
  an unbounded fluid},\ }\href@noop {} {\bibfield  {journal} {\bibinfo
  {journal} {Journal of Fluid Mechanics}\ }\textbf {\bibinfo {volume} {251}},\
  \bibinfo {pages} {411} (\bibinfo {year} {1993})}\BibitemShut {NoStop}%
\bibitem [{\citenamefont {Dassios}\ \emph {et~al.}(1994)\citenamefont
  {Dassios}, \citenamefont {Hadjinicolaou},\ and\ \citenamefont
  {Payatakes}}]{dassios1994generalized}%
  \BibitemOpen
  \bibfield  {author} {\bibinfo {author} {\bibfnamefont {G.}~\bibnamefont
  {Dassios}}, \bibinfo {author} {\bibfnamefont {M.}~\bibnamefont
  {Hadjinicolaou}},\ and\ \bibinfo {author} {\bibfnamefont {A.}~\bibnamefont
  {Payatakes}},\ }\bibfield  {title} {\bibinfo {title} {Generalized
  eigenfunctions and complete semiseparable solutions for stokes flow in
  spheroidal coordinates},\ }\href@noop {} {\bibfield  {journal} {\bibinfo
  {journal} {Quarterly of Applied Mathematics}\ }\textbf {\bibinfo {volume}
  {52}},\ \bibinfo {pages} {157} (\bibinfo {year} {1994})}\BibitemShut
  {NoStop}%
\end{thebibliography}%

\end{document}

\title{Supplementary Material: Shape-induced pairing of spheroidal squirmers}
\author{Ruben Poehnl}
\affiliation{Department of Mechanical Engineering, University of Hawai’i at 
M{\=a}noa, 2540 Dole Street, Holmes Hall 302, Honolulu, Hawaii 96822, USA}

\author{William E. Uspal}
\email{uspal@hawaii.edu}
\affiliation{Department of Mechanical Engineering, University of Hawai’i at 
M{\=a}noa, 2540 Dole Street, Holmes Hall 302, Honolulu, Hawaii 96822, USA}

\date{\today}
\maketitle

\section{Minimal model: collinear bound states \parindent0pt } 
\begin{figure}[h]
    \centering
    \includegraphics[scale=0.6]{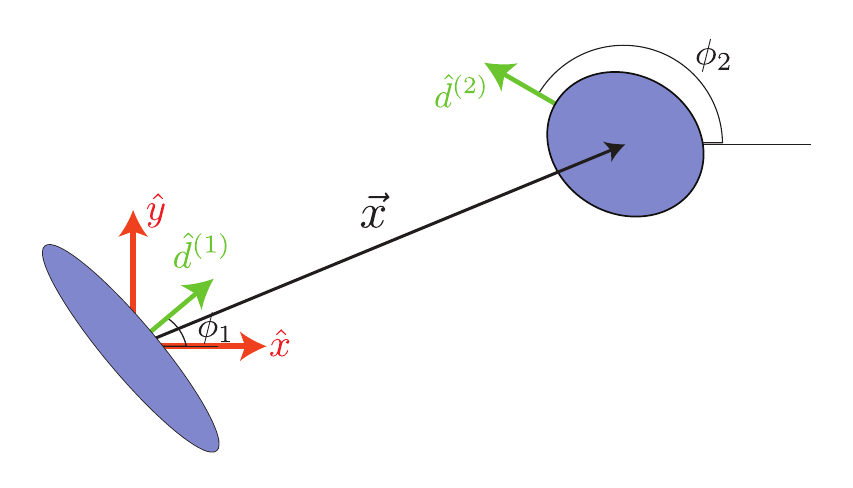}
    \caption{Schematic illustration of the three degrees of freedom $\phi_1$, $\phi_2$, and \textcolor{black}{$d = |\mathbf{x}|$} for a pair of spheroids.}
    \label{fig:two_part_schematic}
\end{figure}

{\parindent0pt For a pair of spheroids,} the three degrees of freedom $\phi_1$, $\phi_2$, and \textcolor{black}{$d = |\mathbf{x}|$} are schematically illustrated in Fig. \ref{fig:two_part_schematic}.
We choose to measure $\phi_1$ and $\phi_2$ with respect to the x-axis in the laboratory frame. \textcolor{black}{Without loss of generality, we choose to place the origin at the instantaneous position of particle 1. Therefore, the position of particle 1 is $\mathbf{x}_1 = (0,0,0)$, and the position of particle 2 is $\mathbf{x}_2 = (x, y, 0)$. In the following, the vector $\mathbf{x} \equiv \mathbf{x}_2 - \mathbf{x}_1$, and $d = |\mathbf{x}| = \sqrt{x^2 + y^2}$.}

\textcolor{black}{Although there are three degrees of freedom, for analytical convenience, we choose to consider the time evolution of four variables: $x$, $y$, $\phi_1$, and $\phi_2$. First, using $\textrm{tr}(\mathbf{S}) = 0$ and Eq. 1 in the main text, we note that
\begin{equation}
\dot{x}_1 = U_s^{(1)} \cos (\phi_1) \; + \frac{3 x  \left(x_i S^{(2)}_{ij} x_j\right)}{8 \pi \mu d^5}
\end{equation}
\begin{equation}
\dot{y}_1 = U_s^{(1)} \sin (\phi_1) + \frac{3 y \left(x_i S^{(2)}_{ij} x_j\right)}{8 \pi \mu d^5} 
\end{equation} 
\begin{equation}
\dot{x}_2 =  U_s^{(2)} \cos (\phi_2) - \frac{3 x  \left(x_i S^{(1)}_{ij} x_j\right)}{8 \pi \mu d^5}
\end{equation}
\begin{equation}
\dot{y}_2 =  U_s^{(2)} \sin (\phi_2)  - \frac{3 y  \left(x_i S^{(1)}_{ij} x_j\right)}{8 \pi \mu d^5},
\end{equation}
where the Einstein summation convention is assumed for repeated subscripts.}
\textcolor{black}{Given that $\dot{x} = \dot{x}_2 - \dot{x}_1$ and $\dot{y} = \dot{y}_2 - \dot{y}_1$, it follows that the  governing equations for the four variables $x$, $y$, $\phi_1$, and $\phi_2$ are as follows:
\begin{equation}
\label{eq:dot_x}
\dot{x} = U_s^{(2)} \cos(\phi_2) -  U_s^{(1)} \cos(\phi_1) - \frac{3 x \left(x_i S_{ij}^{(1)} x_j + x_i S_{ij}^{(2)} x_j\right)  }  {8 \pi \mu (x^2 + y^2)^{5/2}},   
\end{equation}
\begin{equation}
\label{eq:dot_y}
\dot{y} =  U_s^{(2)} \sin(\phi_2) -  U_s^{(1)} \sin(\phi_1) - \frac{3 y \left(x_i S_{ij}^{(1)} x_j + x_i S_{ij}^{(2)} x_j\right)  }  {8 \pi \mu (x^2 + y^2)^{5/2}},      
\end{equation}
\begin{equation}
\dot{\phi_1} = (\hat{d}^{(1)} \times \dot{\hat{d}}^{(1)}) \cdot \hat{z},
\end{equation}
\begin{equation}
\dot{\phi_2} = (\hat{d}^{(2)} \times \dot{\hat{d}}^{(2)}) \cdot \hat{z},
\end{equation}
 Here,
\begin{equation}
  \dot{\hat{d}}^{(1)} = (\mathcal{I} -   \hat{d}^{(1)} \hat{d}^{(1)}) \cdot (\Gamma_1 \mathbf{E}(\mathbf{x}_1)  + \mathbf{W}(\mathbf{x}_1)) \cdot \hat{d}^{(1)}
\end{equation}
\begin{equation}
  \dot{\hat{d}}^{(2)} = (\mathcal{I} -   \hat{d}^{(2)} \hat{d}^{(2)}) \cdot (\Gamma_2 \mathbf{E}(\mathbf{x}_2)  + \mathbf{W}(\mathbf{x}_2)) \cdot \hat{d}^{(2)},
\end{equation}
where $\mathbf{E}(\mathbf{x}_\alpha)$ and $\mathbf{W}(\mathbf{x}_\alpha)$ are the rate-of-strain and vorticity tensors evaluated at the position of particle $\alpha$. These will be given in detail in the next section.  We also recall that 
\begin{equation}
\mathbf{S}^{(1)} = S_{cc}^{(1)}   
 \hat{c}^{(1)} \hat{c}^{(1)} +  S_{dd}^{(1)}   
 \hat{d}^{(1)} \hat{d}^{(1)} + S_{ee}^{(1)}   
 \hat{e}^{(1)} \hat{e}^{(1)}, 
\end{equation}
\begin{equation}
\mathbf{S}^{(2)} = S_{cc}^{(2)}   
 \hat{c}^{(2)} \hat{c}^{(2)} +  S_{dd}^{(2)}   
 \hat{d}^{(2)} \hat{d}^{(2)} + S_{ee}^{(2)}   
 \hat{e}^{(2)} \hat{e}^{(2)}.
\end{equation} 
For the in-plane dynamics, we need the $\hat{c}$, $\hat{d}$, and $\hat{e}$ vectors expressed in terms of $\phi_1$ and $\phi_2$ and the Cartesian unit vectors:
\begin{equation}
\hat{c}^{(1)} = -\sin \phi_1 \, \hat{x} + \cos \phi_1 \, \hat{y},    
\end{equation}
\begin{equation}
\hat{d}^{(1)} = \; \cos \phi_1 \, \hat{x} + \sin \phi_1 \, \hat{y},    
\end{equation}
\begin{equation}
\hat{e}^{(1)} = 0,    
\end{equation}
as well as the corresponding expressions for $\hat{c}^{(2)}$, $\hat{d}^{(2)}$, and $\hat{e}^{(2)}$.
We thus obtain 
\begin{align}
\mathbf{S}^{(1)} = S_{cc}^{(1)} \left[\sin^2 \phi_1 \, \hat{x} \hat{x} - \sin \phi_1 \cos \phi_1 \, (\hat{x} \hat{y} + \hat{y} \hat{x} ) + \cos^2 \phi_1 \, \hat{y}\hat{y}   \right] \\ +\; S_{dd}^{(1)} \left[ \cos^2 \phi_1 \, \hat{x} \hat{x} + \sin \phi_1 \cos \phi_1 (\hat{x} \hat{y} + \hat{y} \hat{x}) + \sin^2 \phi_1 \, \hat{y} \hat{y} \right], \nonumber
\end{align}
and the corresponding expression for $\mathbf{S}^{(2)}$. These expressions are convenient for implementation in computer algebra, along with 
\begin{equation}
\dot{x} =  U_s^{(2)} \cos(\phi_2) -  U_s^{(1)} \cos(\phi_1) - \frac{3 x \left(\mathbf{x} \cdot \mathbf{S}^{(1)} \cdot \mathbf{x} + \mathbf{x} \cdot \mathbf{S}^{(2)} \cdot \mathbf{x} \right)  }  {8 \pi \mu (x^2 + y^2)^{5/2}},   
\end{equation}
\begin{equation}
\dot{y} =   U_s^{(2)} \sin(\phi_2) -  U_s^{(1)} \sin(\phi_1) - \frac{3 y \left(\mathbf{x} \cdot \mathbf{S}^{(1)} \cdot \mathbf{x} + \mathbf{x} \cdot \mathbf{S}^{(2)} \cdot \mathbf{x} \right)  }  {8 \pi \mu (x^2 + y^2)^{5/2}}.
\end{equation}} 
{\parindent0pt \textcolor{black}{\subsection{Head-to-head bound state}  }}
\parindent0pt \textcolor{black}{ First, we consider the head-to-head dynamics ($x \neq 0 $, $y = 0$, $\phi_1 = 0$, and $\phi_2 = \pi$). By symmetry, $\dot{y} = 0$, $\dot{\phi}_1 = 0$, and $\dot{\phi}_2 = 0$. Moreover, $\vec{x} = x \hat{x}$. It follows that
\begin{equation}
\mathbf{S}^{(1)} = S_{cc}^{(1)} \hat{y} \hat{y} + S_{dd}^{(1)} \hat{x} \hat{x},     
\end{equation}
\begin{equation}
\mathbf{S}^{(2)} = S_{cc}^{(2)} \hat{y} \hat{y} + S_{dd}^{(2)} \hat{x} \hat{x},     
\end{equation}
and therefore
\begin{equation}
x_i S_{ij}^{(1)} x_j = x^2 S_{dd}^{(1)},   
\end{equation}
\begin{equation}
x_i S_{ij}^{(2)} x_j = x^2 S_{dd}^{(2)}.
\end{equation}
The equation of motion for $x$ becomes
\begin{equation}
\dot{x} = -U_s^{(2)} - U_s^{(1)} - \frac{3(S_{dd}^{(1)} + S_{dd}^{(2)})}{8 \pi \mu x^2 }.    
\end{equation}
Solving for $\dot{x} = 0$ and $x = d_0$ gives the head-to-head bound state separation
\begin{equation}
d_0 = \sqrt{\frac{-3\left(S_{dd}^{(1)} + S_{dd}^{(2)}\right)}{8 \pi \mu (U_s^{(2)} + U_s^{(1)})}}.  
\end{equation}
}
{\textcolor{black}{\parindent0pt \subsection{Head-to-tail bound state}}}
\textcolor{black}{For a head-to-tail configuration ($x \neq 0$, $y = 0$, $\phi_1 = 0$, and $\phi_2 = 0$), the equation of motion for $x$ is identical, except for the sign of the first term:
\begin{equation}
\dot{x} = U_s^{(2)} - U_s^{(1)} - \frac{3 \left(S_{dd}^{(1)} + S_{dd}^{(2)}\right)}{8 \pi \mu x^2 }.
\end{equation}
We therefore obtain
\begin{equation}
d_0 = \sqrt{\frac{3 \left(S_{dd}^{(1)} + S_{dd}^{(2)}\right)}{8 \pi \mu (U_s^{(2)} - U_s^{(1)})}}.  
\end{equation}
Both particles move with the same steady velocity. We can obtain the velocity $U$ of the pair by calculating the velocity of particle 1:
\begin{equation}
U = U_s^{(1)} + \frac{3 S^{(2)}_{dd}}{8 \pi \mu d_0^2},
\end{equation}
\begin{equation}
U = U_s^{(1)} + \frac{S_{\textcolor{black}{dd}}^{(2)} (U_s^{(2)} - U_s^{(1)})}{S_{\textcolor{black}{dd}}^{(1)} + S_{\textcolor{black}{dd}}^{(2)}}.
\end{equation}
}
\section{Minimal model: linear stability analysis \parindent0pt} 
{\parindent0pt \textcolor{black}{The linear stability of a bound state (i.e., fixed point) can be determined from the eigenvalues of the Jacobian.}} For a fixed point, the Jacobian is constructed as follows:
\begin{equation}
\mathbf{J} = \begin{pmatrix}
\frac{\partial \dot{x}}{\partial x} & \frac{\partial \dot{x}}{\partial y} & \frac{\partial \dot{x}}{\partial \phi_1} & \frac{\partial \dot{x}}{\partial \phi_2} \\
\frac{\partial \dot{y}}{\partial x} & \frac{\partial \dot{y}}{\partial y} & \frac{\partial \dot{y}}{\partial \phi_1} & \frac{\partial \dot{y}}{\partial \phi_2} \\

\frac{\partial \dot{\phi_1}}{\partial x} & \frac{\partial \dot{\phi_1}}{\partial y} & \frac{\partial \dot{\phi_1}}{\partial \phi_1} & \frac{\partial \dot{\phi_1}}{\partial \phi_2} \\

\frac{\partial \dot{\phi_2}}{\partial x} & \frac{\partial \dot{\phi_2}}{\partial y} & \frac{\partial \dot{\phi_2}}{\partial \phi_1} & \frac{\partial \dot{\phi_2}}{\partial \phi_2} \\

\end{pmatrix}_{\textcolor{black}{x=d_0},y=0,\phi_1 = 0, \textcolor{black}{\phi_2 = 0\;\textrm{or}\;\phi_2=\pi}}
\end{equation}
\textcolor{black}{where the choice of $\phi_2 = 0$ or $\phi_2 = \pi$ corresponds to a head-to-tail or head-to-head configuration, respectively.}
The Jacobian can be obtained numerically, by approximating the partial derivatives using finite differences,  or analytically.

\textcolor{black}{At this point, it is necessary to specify $\mathbf{E}(\mathbf{x}_\alpha)$ and $\mathbf{W}(\mathbf{x}_\alpha)$ as functions of $x$, $y$, $\phi_1$, and $\phi_2$. The rate-of-strain and vorticity tensors can be constructed from the fluid velocity field: 
\begin{equation}
\mathbf{E} = \frac{1}{2} \left(\nabla \mathbf{u} + \nabla \mathbf{u}^T \right),    
\end{equation}
\begin{equation}
\mathbf{W} = \frac{1}{2} \left(\nabla \mathbf{u} - \nabla \mathbf{u}^T \right).    
\end{equation}
Using $\textrm{tr}(\mathbf{S}) = 0$ and Eq. 4 in the main text, one can  obtain the following partial derivative of the velocity field due to a stresslet located at the origin:
\begin{equation}
\frac{\partial u_i}{\partial x_\ell} =  \frac{1}{8 \pi \mu r^5} \left( -3 x_j x_k \delta_{i \ell} S_{jk} 
 - 3 x_i x_k S_{\ell k} - 3 x_i x_j S_{j \ell} + \frac{15 x_i x_j x_k x_\ell S_{jk}}{r^2} \right).
\end{equation}
Using $S_{j \ell} = S_{\ell j}$, we can combine the second and third terms in the parentheses:
\begin{equation}
\frac{\partial u_i}{\partial x_\ell} =  \frac{1}{8 \pi \mu r^5} \left( -3 x_j x_k \delta_{i \ell} S_{jk} 
 - 6 x_i x_k S_{\ell k} + \frac{15 x_i x_j x_k x_\ell S_{jk}}{r^2} \right).
\end{equation}
The rate-of-strain tensor is 
\begin{equation}
E_{i \ell} = \frac{1}{8 \pi \mu r^5} \left(-3 x_j x_k \delta_{i \ell}  S_{jk} - 3 x_i x_k S_{\ell k} - 3 x_\ell x_k S_{ik} +  \frac{15 x_i x_j x_k x_\ell S_{jk}}{r^2}  \right).
\end{equation}
For implementation in computer algebra, it is convenient to express this quantity in vector notation:
\begin{equation}
\label{eq:Evecnot}
\mathbf{E} = \frac{1}{8 \pi \mu r^5} \left( -3 \, \mathcal{I} \, (\mathbf{x} \cdot \mathbf{S} \cdot \mathbf{x}) - 3 \, \mathbf{x} \; (\mathbf{S} \cdot \mathbf{x}) - 3 \, (\mathbf{S} \cdot \mathbf{x}) \; \mathbf{x} + \frac{15 (\mathbf{x} \cdot \mathbf{S} \cdot \mathbf{x}) \; \mathbf{x} \mathbf{x}}{r^2}  \right). 
\end{equation}
Concerning the vorticity tensor $W_{ij}$, it is important to note that we are using the convention $(\nabla \mathbf{u})_{ij} \equiv \frac{\partial u_i}{\partial x_j}$.  Thus, $W_{ij} = \frac{1}{2} \left(\frac{\partial u_i}{\partial u_j} - \frac{\partial u_j}{\partial u_i}\right)$. Since the vorticity tensor is anti-symmetric, its sign is sensitive to the choice of convention. (The consistency of this convention with the Jeffery equation as given in Eq. 2 of the main text can be checked by noting that the vorticity contribution $\left(\mathcal{I} -  \hat{d}^{(\alpha)}  \hat{d}^{(\alpha)} \right) \cdot \mathbf{W} \cdot \hat{d}^{(\alpha)}$ should give a contribution identical to  $\frac{1}{2} \bm{\omega} \times \hat{d}^{(\alpha)}$, where $\bm{\omega} = \nabla \times \mathbf{u}$.) We obtain 
\begin{equation}
W_{i \ell} = \frac{3}{8 \pi \mu r^5} \left( -x_i x_k S_{\ell k} + x_\ell x_k S_{i k} \right),
\end{equation}
or
\begin{equation}
\label{eq:Wvecnot}
\mathbf{W} = \frac{3}{8 \pi \mu r^5}  \left( -\mathbf{x} \; (\mathbf{S} \cdot \mathbf{x}) \;  +  (\mathbf{S} \cdot \mathbf{x}) \; \mathbf{x} \right).
\end{equation}
We note that only even numbers of factors of $\mathbf{x}$ appear in each term in Eq. \ref{eq:Evecnot} and Eq. \ref{eq:Wvecnot}. Therefore, no sign inversion is needed to use these expressions to find the flow components at particle 1 induced by particle 2.  This results from the nematic symmetry of the stresslet. It follows that 
\begin{equation}
\mathbf{E}(\mathbf{x}_1) =  \frac{1}{8 \pi \mu d^5} \left( -3 \, \mathcal{I} \, (\mathbf{x} \cdot \mathbf{S}^{(2)} \cdot \mathbf{x}) - 3 \, \mathbf{x} \; (\mathbf{S}^{(2)} \cdot \mathbf{x}) - 3 \, (\mathbf{S}^{(2)} \cdot \mathbf{x}) \; \mathbf{x} + \frac{15 (\mathbf{x} \cdot \mathbf{S}^{(2)} \cdot \mathbf{x}) \; \mathbf{x} \mathbf{x}}{d^2}  \right),
\end{equation}
\begin{equation}
\mathbf{W}(\mathbf{x}_1) = \frac{3}{8 \pi \mu d^5}  \left( -\mathbf{x} \; (\mathbf{S}^{(2)} \cdot \mathbf{x}) \;  +  (\mathbf{S}^{(2)} \cdot \mathbf{x}) \; \mathbf{x} \right)    ,
\end{equation}
and corresponding expressions for particle 2, where $\mathbf{x}$ is again the vector from particle 1 to particle 2, and $d = |\mathbf{x}| = \sqrt{x^2 + y^2}$.
}

{\parindent0pt \subsection{Head-to-head bound state}}
{\parindent0pt From the governing dynamical system,} we analytically \textcolor{black}{(with computer algebra)} obtain the Jacobian
\begin{equation}
\mathbf{J} = \begin{pmatrix} 
\frac{3 \left(S_{dd}^{(1)} + S_{dd}^{(2)} \right)}{4 \pi \mu \textcolor{black}{d_0}^3} & 0 & 0 & 0 \\

0 & - \frac{3 \left(S_{dd}^{(1)} + S_{dd}^{(2)} \right)}{8 \pi \mu \textcolor{black}{d_0}^3} & -U_{s}^{(1)} & -U_{s}^{(2)}\\

0 & -\frac{3 (S_{dd}^{(2)} (1- 4 \Gamma_1) + S_{cc}^{(2)} (-1 + \Gamma_1))}{8 \pi \mu \textcolor{black}{d_0}^4} & -\frac{9 S_{dd}^{(2)} \Gamma_1}{8 \pi \mu \textcolor{black}{d_0}^3} & \frac{3 (S_{cc}^{(2)} - S_{dd}^{(2)})(-1 + \Gamma_1)}{8 \pi \mu \textcolor{black}{d_0}^3} \\

0 & -\frac{3 (S_{dd}^{(1)} (1- 4 \Gamma_2) + S_{cc}^{(1)} (-1 + \Gamma_2))}{8 \pi \mu \textcolor{black}{d_0}^4}  & \frac{3 (S_{cc}^{(1)} - S_{dd}^{(1)})(-1 + \Gamma_2)}{8 \pi \mu \textcolor{black}{d_0}^3} & -\frac{9 S_{dd}^{(1)} \Gamma_2}{8 \pi \mu \textcolor{black}{d_0}^3} 
\end{pmatrix}.
\end{equation}
\textcolor{black}{From the zeroes in the first row and first column, we note that translations in the $x$ direction decouple from translations in $y$ and rotations. By inspection, we immediately} obtain the first requirement for stability: $(S_{dd}^{(1)} + S_{dd}^{(2)}) < 0$. \textcolor{black}{This is the ``net pusher'' requirement.} Next, we consider the densely populated 3x3 submatrix obtained by removing the first row and first column of $\mathbf{J}$. The characteristic equation of this submatrix is:
\begin{equation}
\lambda^3 + A \lambda^2 + B \lambda = 0,
\end{equation}
with 
\begin{equation}
A = \frac{3 (S_{dd}^{(1)} + S_{dd}^{(2)}) + 9 S_{dd}^{(2)} \Gamma_1 + 9 S_{dd}^{(1)} \Gamma_2 }{8 \pi \mu \textcolor{black}{d_0}^3}
\end{equation}
and
\begin{align}
\label{eq:B_head_to_head}
B = \frac{-3 U_s^{(1)} (S_{dd}^{(2)} (1 - 4 \Gamma_1) + S_{cc}^{(2)} (-1 + \Gamma_1) ) - 3 U_s^{(2)}(S_{dd}^{(1)}(1 - 4 \Gamma_2) + S_{cc}^{(1)}(-1 + \Gamma_2)) }{8 \pi \mu \textcolor{black}{d_0}^4} \; + \nonumber \\ 
\frac{27 (S_{dd}^{(2)} \Gamma_1 + S_{dd}^{(1)} \Gamma_2) (S_{dd}^{(1)} + S_{dd}^{(2)}) + 81 S_{dd}^{(1)} S_{dd}^{(2)}    \Gamma_1 \Gamma_2  - 9 (S_{cc}^{(1)} - S_{dd}^{(1)}) (S_{cc}^{(2)} - S_{dd}^{(2)}) (-1 + \Gamma_1) (-1 + \Gamma_2) }{64 \pi^2 \mu^2 \textcolor{black}{d_0}^6}.
\end{align}
It is immediately apparent that one of the eigenvalues is $\lambda = 0$. This is to be expected, since, for convenience, we considered more coordinates (four) than was needed to specify the state of the system (three). The $\lambda = 0$ eigenvalue therefore corresponds to a symmetry of the dynamical system. Factoring out this eigenvalue, we obtain
\begin{equation}
\label{eq:characteristic_quadratic}
\lambda^2 + A \lambda + B = 0.
\end{equation}
Although we could solve for $\lambda$, we instead apply the Routh-Hurwitz stability criterion to determine the two remaining requirements for dynamical stability. For a quadratic characteristic equation,  the Routh-Hurwitz criterion specifies that a necessary and sufficient condition for stability is that all coefficients of the equation are positive. In other words, we require $A > 0$ and $B > 0$. 
The requirement on $A$ gives
\begin{equation}
(S_{dd}^{(1)} + S_{dd}^{(2)}) + 3(S_{dd}^{(2)} \Gamma_1 +  S_{dd}^{(1)} \Gamma_2) > 0.
\end{equation}
For a pair of identical particles, the requirement $A > 0$ reduces to $\Gamma < -1/3$, and the requirement $B > 0$ reduces to $[S_{cc}(-1 + \Gamma) + S_{dd}(1 + 2 \Gamma)] [S_{cc} (-1 + \Gamma) - S_{dd}(1 + 4 \Gamma)] < 0$. For a pair of identical \textit{axisymmetric} squirmers, the requirement $B > 0$ further reduces to $(1 + \Gamma) (1 + 9 \Gamma) > 0$. Given that $-1 < \Gamma < 1$ by definition, this requirement finally reduces to $\Gamma > -1/9$. Thus, for axisymmetric squirmers, we obtain incompatible requirements.
\newline

{\parindent0pt \subsection{Head-to-tail bound state}}
The Jacobian matrix for the head-to-tail bound state is identical to the Jacobian for the head-to-head bound state, except that the entry in the second row, fourth column in $U_s^{(2)}$. Thus, the first requirement for stability is again $(S_{dd}^{(1)} + S_{dd}^{(2)}) < 0$, \textcolor{black}{i.e., the pair is a ``net pusher.''} As before, the characteristic equation reduces to Eq. \ref{eq:characteristic_quadratic}. Moreover, the coefficient $A$ is again
\begin{equation}
A = \frac{3 (S_{dd}^{(1)} + S_{dd}^{(2)}) + 9 S_{dd}^{(2)} \Gamma_1 + 9 S_{dd}^{(1)} \Gamma_2 }{8 \pi \mu \textcolor{black}{d_0}^3},
\end{equation}
giving the condition 
\begin{equation}
(S_{dd}^{(1)} + S_{dd}^{(2)}) + 3(S_{dd}^{(2)} \Gamma_1 +  S_{dd}^{(1)} \Gamma_2) > 0.
\label{eq:ht_criterion1}
\end{equation}
The coefficient $B$ is nearly identical to Eq. \ref{eq:B_head_to_head}, but differs in the sign of the second term in the numerator of the fraction with $d^4$ in the denominator:
\begin{equation}
\begin{split}
B = \frac{-3 U_s^{(1)} (S_{dd}^{(2)} (1 - 4 \Gamma_1) + S_{cc}^{(2)} (-1 + \Gamma_1) ) + 3 U_s^{(2)}(S_{dd}^{(1)}(1 - 4 \Gamma_2) + S_{cc}^{(1)}(-1 + \Gamma_2)) }{8 \pi \mu \textcolor{black}{d_0}^4} \; + \\ 
\frac{27 (S_{dd}^{(2)} \Gamma_1 + S_{dd}^{(1)} \Gamma_2) (S_{dd}^{(1)} + S_{dd}^{(2)}) + 81 S_{dd}^{(1)} S_{dd}^{(2)}    \Gamma_1 \Gamma_2  - 9 (S_{cc}^{(1)} - S_{dd}^{(1)}) (S_{cc}^{(2)} - S_{dd}^{(2)}) (-1 + \Gamma_1) (-1 + \Gamma_2) }{64 \pi^2 \mu^2 \textcolor{black}{d_0}^6}.
\end{split}
\end{equation}
We recall that the second condition for stability is $B > 0$.
\\
{\parindent0pt \subsubsection{Head-to-tail bound states of axisymmetric squirmers.}}
{\parindent0pt For non-identical \textit{axisymmetric} squirmers,} the condition $A > 0$ becomes 
\begin{equation}
\sigma_0^{(1)} +   \sigma_0^{(2)} + 3 (\sigma_0^{(2)} \Gamma_1 + \sigma_0^{(1)} \Gamma_2) > 0.
\end{equation}
The condition $B > 0$ becomes, using the ``net pusher'' criterion \textcolor{black}{$\sigma_0^{(1)} + \sigma_0^{(2)} < 0$},
%\begin{equation}
%\begin{split}
%2 \sigma_0^{(1)}^2 \left(U_s^{(2)}-\Gamma_2 \left(2 U_s^{(1)}+U_s^{(2)}\right)\right) \\
%+\sigma _0^{(2)} \sigma _0^{(1)} \left[\left( \Gamma_2+ \left(9 \Gamma _2+7\right) \Gamma_1 -1\right) U_{s}^{(2)}
%-\left(\Gamma_1+\left(9 \Gamma _1+7\right) \Gamma_2-1\right) U_s^{(1)}\right] \\ +2 \sigma _0^{(2)}^2 \left(\left(\Gamma _1-1\right) U_{s}^{(1)}+2 \Gamma _1 U_{{s}}^{(2)}\right)  < 0.
%\end{split}
%\end{equation}
%\begin{equation}
%\begin{split}
%2 \sigma_0^{(1)}^2 \left(U_s^{(2)}-\Gamma_2 \left(2 U_s^{(1)}+U_s^{(2)}\right)\right) \\
%+\sigma _0^{(2)} \sigma _0^{(1)} \left[\left( \Gamma_2+ \left(9 \Gamma _2+7\right) \Gamma_1 -1\right) U_{s}^{(2)}
%-\left(\Gamma_1+\left(9 \Gamma _1+7\right) \Gamma_2-1\right) U_s^{(1)}\right] \\  -2 \sigma _0^{(2)}^2 \left(U_{s}^{(1)}-\Gamma _1 (2 U_{{s}}^{(2)}+U_{s}^{(1)})\right)  < 0.
%\end{split}
%\end{equation}
\begin{equation}
\begin{split}
        \left(-2 \left(\Gamma _2-1\right) \sigma _1^2+\left(\Gamma _2+\Gamma _1 \left(9 \Gamma _2+7\right)-1\right) \sigma_0 ^{(2)} \sigma_0^{(1)}+4 \Gamma _1 {\sigma_0^{(2)}}^2\right) U_{\text{s}}^{(2)} \\ - \left(\Gamma _2 \sigma_0^{(1)} \left(\left(9 \Gamma _1+7\right) \sigma_0^{(2)}+4 \sigma_0^{(1)}\right)+\left(\Gamma _1-1\right) \sigma_0^{(2)} \left(\sigma _0^{(1)}-2 \sigma _0^{(2)}\right)\right) U_{\text{s}}^{(1)} < 0.
        \end{split}
\end{equation}
Defining the parameters $S \equiv \sigma^{(2)}_0/\sigma^{(1)}_0$ and $V \equiv U_s^{2}/U_s^{1}$, with $V \geq 0$, we obtain
\begin{equation}
\label{eq:S_A_criterion}
\textrm{sgn}\left(\sigma_0^{(1)}\right) \, [1  +   S + 3 (S \Gamma_1 + \Gamma_2)] > 0
\end{equation}
and
\begin{equation}
\label{eq:B_criterion_V_S}
\begin{split}
\Gamma _2 ((S-2) V-7 S-4)+\Gamma _1 S \left(4 S V+2 S+9 \Gamma _2 (V-1)+7 V-1\right) \\ 
-S (2 S+V-1)+2 V  < 0.
\end{split}
\end{equation}
In addition, the ``net pusher'' criterion can be expressed in term of $S$:
\begin{equation}
\label{eq:net_pusher_S_criterion}
\textrm{sgn}\left(\sigma_0^{(1)}\right) (1 + S) < 0. 
\end{equation}
These expressions make clear that the phase boundary exists in a space defined by four parameters: $\Gamma_1$, $\Gamma_2$, $S$, and $V$. 

It is instructive to fix $\Gamma_1$,  $\Gamma_2$, and $\textrm{sgn}\left(\sigma_0^{(1)}\right)$, and consider the two-dimensional phase subspace defined by $V$ and $S$. First, we note that Eqs. \ref{eq:S_A_criterion} and \ref{eq:net_pusher_S_criterion} do not involve $V$, and place numerical bounds on $S$. Then, considering Eq. \ref{eq:B_criterion_V_S}, we note that it can be written in the form 
\begin{equation}
\label{eq:VeqnLinear}
f_1(S) + f_2(S) V < 0,    
\end{equation}
where $f_1(S)$ and $f_2(S)$ are functions of $S$. Depending on the numerical bounds on $S$, satisfaction of the condition given in Eq. \ref{eq:VeqnLinear} may or may not depend on the value of $V$. As a concrete example, we consider $\textrm{sgn}\left(\sigma_0^{(1)}\right) = -1$, $\Gamma_1 = -0.7$, and $\Gamma_2 = 0.2$. From Eq.  \ref{eq:net_pusher_S_criterion}, we obtain $S > -1$. From Eq. \ref{eq:S_A_criterion}, we obtain a stricter lower bound, $S > 1.46$. Eq. \ref{eq:B_criterion_V_S} gives $(-0.8 + 1.56 S - 3.4 S^2) + (1.6 - 6.96 S - 2.8 S^2) V < 0$. Both terms in parentheses are negative for $S > 1.46$. Therefore, the condition is satisfied for all $V > 0$, and the phase boundary reduces to a line $S = 1.46$ in $(V,S)$ space.

%Fig. XX shows phase boundaries in $(V,S)$ for selected values of $(\Gamma_1, \Gamma_2)$, with $\textrm{sgn}\left(\sigma_0^{(1)}\right) = -1$.)

\vspace{0.12in}
{\parindent0pt \subsubsection{A no-cargo theorem for far-field bound states.}} \textcolor{black}{Using active colloids as ``cargo carriers'' is often suggested as a desirable potential application in lab-on-a-chip systems.} {\parindent0pt In the framework of  \textcolor{black}{our far-field theory, we now show that}} it is not possible to have a stable \textcolor{black}{collinear} bound state in which one of the particles is an inert cargo with no self-propulsion and no active stresslet. First, given the criterion $U_s^{(1)} > U_s^{(2)}$, it is clear that the trailing particle cannot be an inert cargo. Now we consider the possibility that particle 2 is inert, i.e., that $U_s^{(2)} = 0$,  $S_{cc}^{(2)} = 0$, $S_{dd}^{(2)} = 0$, and $S_{ee}^{(2)} = 0$. Eq. \ref{eq:ht_criterion1} gives:
\begin{equation}
    S_{dd}^{(1)}(1 + 3 \Gamma_2) > 0.
\end{equation}
Since the pair must be  a net squirmer, $S_{dd}^{(1)} < 0$, so that $\Gamma_2 < -1/3$. Similarly, using $S_{dd}^{(1)} < 0$, the criterion $B > 0$ yields
$\Gamma_2 > 0$.
We therefore obtain a contradiction.  \textcolor{black}{We note that this result does not exclude the possibility of ``cargo-carrying'' sustained by near-field hydrodynamics or lubrication forces, particle or fluid inertia, phoretic/chemical interactions, or other effects not included in our minimal theory.}\newline

\section{Properties of the stresslet tensor}

\textcolor{black}{Eq. 6 in the main text is generic because the stresslet tensor is symmetric and real-valued. Thus, the principal axes are orthogonal, and one can always rotate to a frame in which the stresslet tensor is diagonal.}

\textcolor{black}{However, we note some technical caveats. First, the particle’s propulsion velocity may or may not be aligned with one of the principal axes. Subsequent to Eq. 6 in the main text, we assume that the propulsion velocity is aligned with a principal axis $\hat{d}$.  Secondly, the principal axes may or may not be aligned with axes of a Cartesian coordinate system that naturally corresponds to a confining geometry. 
We can give an example in which both of these complications occur. We consider a discoidal ICEP particle near a solid planar wall, as in Katuri \textit{et al}. \cite{katuri22} The AC field is oriented normal to the wall. The particle’s axis of (geometric and material) symmetry is oriented parallel to the wall, and the particle center has height h above the wall. As the most natural coordinate system for the ``laboratory'' frame, we define the z-direction as being aligned with the wall normal, and the x- and y-directions to lie in the plane of the wall. In this frame, we numerically calculate, using Eq. 1 in the main text and theoretical methods in Katuri \textit{et al}., that the stresslet tensor is not diagonal. (Two off-diagonal elements are roughly 3x smaller in magnitude than the smallest magnitude element on the diagonal.) Additionally, we find that the particle velocity vector is not aligned with a principal axis of the stresslet.  
However, when the particle is far from the wall (h is large), the off-diagonal components of the stresslet tensor diminish, and the particle’s direction of propulsion is nearly aligned with one of the principal axes. Thus, the analysis in the present work is a good approximation for ICEP particles in free space.
It would be relatively straightforward to extend the analysis to accommodate the effects mentioned above. This could be considered in future work that is oriented towards application to ICEP particles moving near a planar substrate. In the present case, these technicalities are not considered in order to maintain a simple and clear presentation.}

\section{Spheroidal coordinate systems}
\textcolor{black}{To test our analytical predictions, we compare the results to numeric calculations with the boundary element method. In its implementation, we use prolate and oblate spheroidal coordinates. In this Section, we refer to the Cartesian coordinate system in Fig. 1 of the main text, but omit the prime symbols for notational clarity. The prolate spheroidal coordinates $(\tau,\xi,\varphi)$ are related to the Cartesian coordinates via the transformation \cite{dassios1994generalized}
\begin{align}
\label{eq:prol_coord}
x&=c\sqrt{(\tau^2-1)(1-\xi^2)}\sin(\varphi)\,,\nonumber\\
y&=c\tau\xi \,,\nonumber\\
z&=c\sqrt{(\tau^2-1)(1-\xi^2)}\cos(\varphi)\,,\nonumber
\end{align}
with the geometric prefactor $c=\sqrt{1-r_e^2}$ and have basis vectors 
\begin{eqnarray}
 \label{eq:unit_vec}
\bold{e}_\tau &=& \left(\frac{\tau \cdot \sqrt{1-\xi^2}}{\sqrt{\tau^2-1}} 
\bold{e}_z + \xi \bold{e}_y \right) \cdot 
\frac{\sqrt{\tau^2-1}}{\sqrt{\tau^2-\xi^2}}\,, \nonumber\\
\bold{e}_\xi &=& \left (\frac{\xi \cdot 
\sqrt{\tau^2-1}} {\sqrt{1-\xi^2}} \bold{e}_z + 
\tau \bold{e}_y \right) \cdot \frac{\sqrt{1-\xi^2}}{\sqrt{\tau^2-\xi^2}}\,
\end{eqnarray}
for $\varphi=0$.
Similarly, the transformations for the oblate spheroidal coordinates $(\lambda,\xi,\varphi)$
\begin{align}
x&=\bar{c}\sqrt{(1+\lambda^2)(1-\xi^2)}\sin(\varphi)\,,\nonumber\\
y&=\bar{c}\lambda\xi \,,\nonumber\\
z&=\bar{c}\sqrt{(1+\lambda^2)(1-\xi^2)}\cos(\varphi)\,,\nonumber
\end{align}
use a geometric prefactor $\bar{c}=\sqrt{r_e^2-1}$ and have basis vectors
\begin{eqnarray}
 \label{eq:unit_vec}
\bold{e}_\lambda &=& \left(\frac{\lambda \cdot \sqrt{1-\xi^2}}{\sqrt{1+\lambda^2}} 
\bold{e}_z + \xi \bold{e}_y \right) \cdot 
\frac{\sqrt{1+\lambda^2}}{\sqrt{\lambda^2+\xi^2}}\,, \nonumber\\
\bold{e}_\xi &=& \left (\frac{-\xi \cdot 
\sqrt{1+\lambda^2}} {\sqrt{1-\xi^2}} \bold{e}_z + 
\lambda \bold{e}_y \right) \cdot \frac{\sqrt{1-\xi^2}}{\sqrt{\lambda^2+\xi^2}}\,
\end{eqnarray}
for $\varphi=0$.}
\section{Velocity and stresslet of axisymmetric spheroidal squirmer}
\begin{figure}
\centering
(a) {\includegraphics[width=0.35\textwidth]{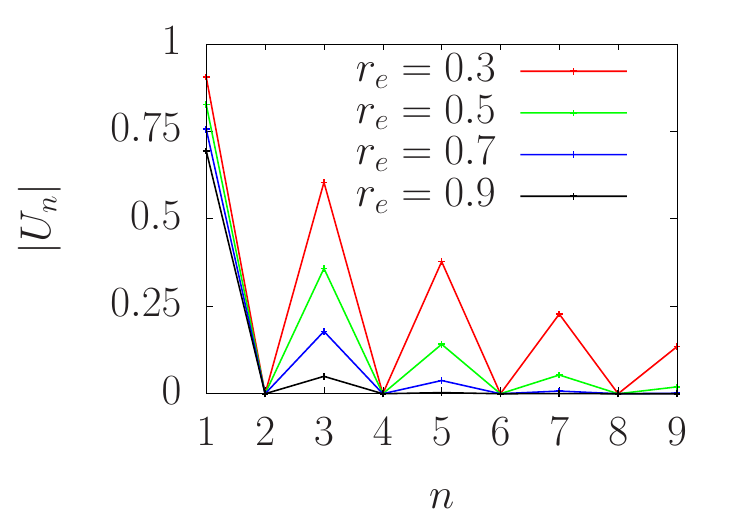}}
(b) {\includegraphics[width=0.35\textwidth]{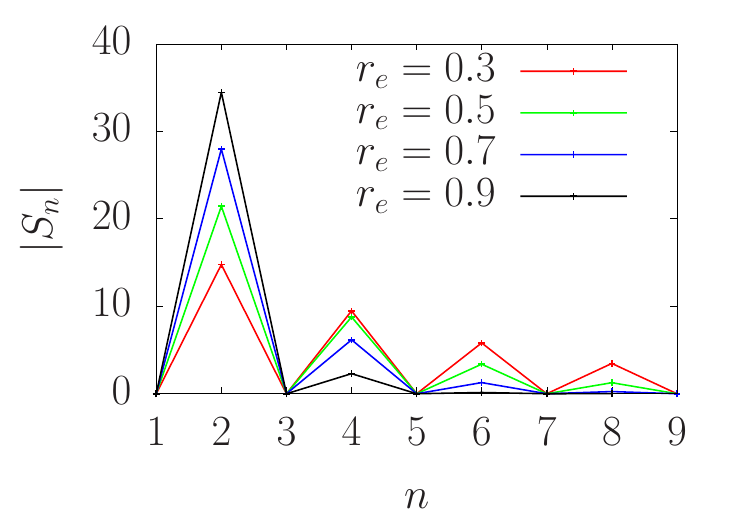}}
\newline
(c) {\includegraphics[width=0.35\textwidth]{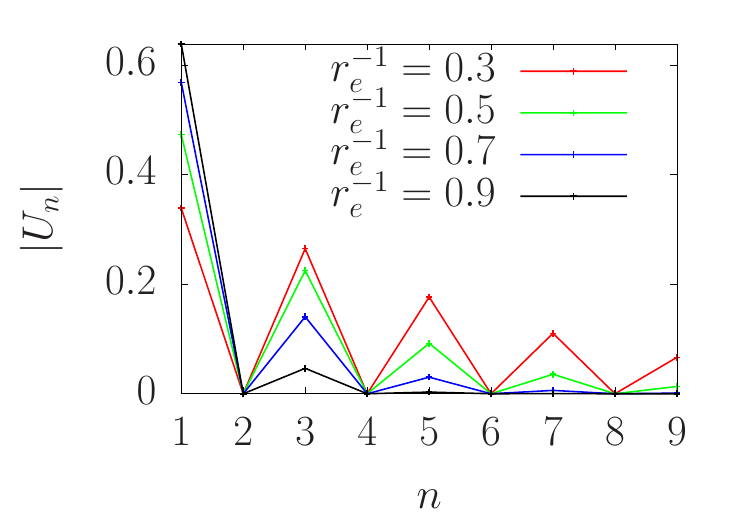}}
(d) {\includegraphics[width=0.35\textwidth]{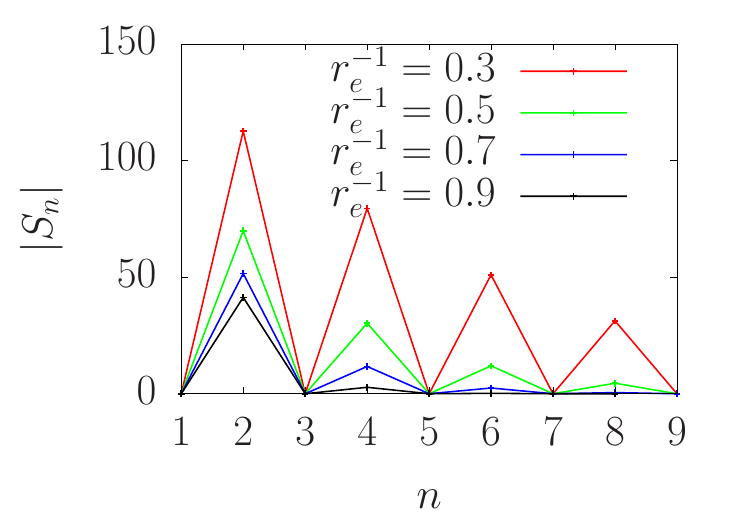}}
\caption{(a) The absolute value $|U_n|$ of the contribution of the slip modes $n = 1,3,5,7$ to the velocity of a prolate squirmer for aspect ratios $r_e=0.3,0.5,0.7,0.9$. (b) The absolute value $|S_n|$ of the contribution of the slip modes $n = 1,3,5,7$ to the stresslet of a prolate squirmer for aspect ratios $r_e=0.3,0.5,0.7,0.9$. (c) The absolute value $|U_n|$ of the contribution of the slip modes $n = 1,3,5,7$ to the velocity of an oblate squirmer for aspect ratios $r_e^{-1}=0.3,0.5,0.7,0.9$. (d) The absolute value $|S_n|$ of the contribution of the slip modes $n = 1,3,5,7$ to the stresslet of an oblate squirmer for aspect ratios $r_e^{-1}=0.3,0.5,0.7,0.9$. The lines represent only a guide to the eye.}
\label{Fig:corr_velo+stress}
\end{figure}

\begin{figure}
\centering
(a) {\includegraphics[width=0.35\textwidth]{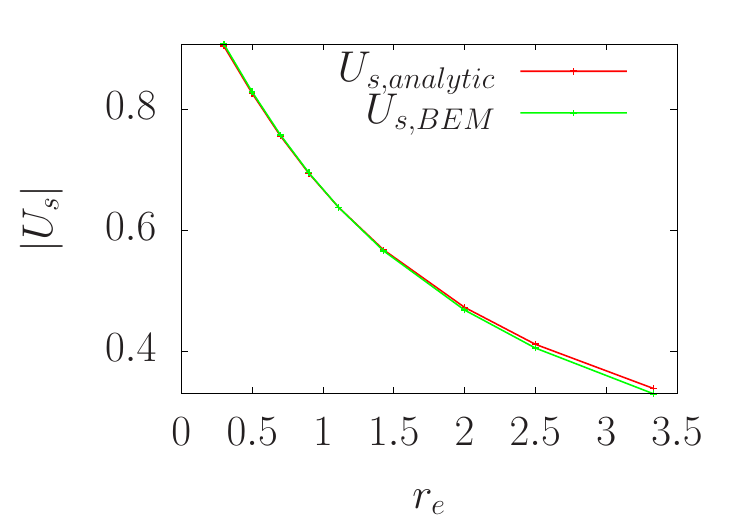}}
(b)
 {\includegraphics[width=0.35\textwidth]{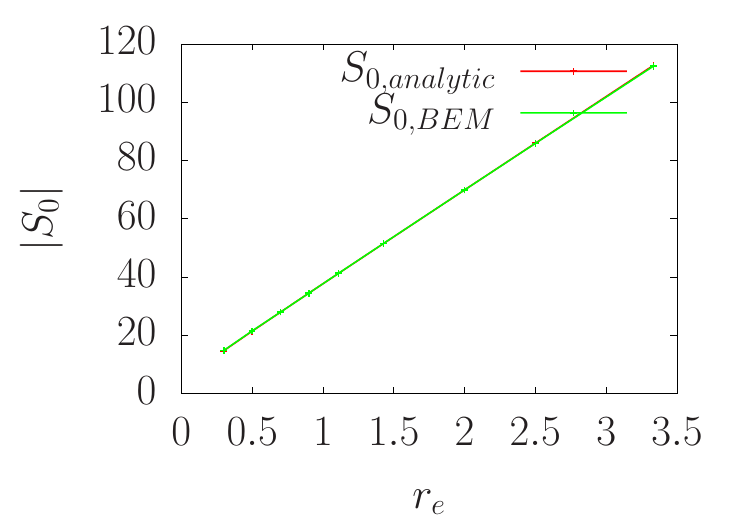}}
\caption{  (a) Comparison between the analytically predicted and the numerically calculated velocity $|U_s|$ over a range of aspect ratios $r_e$. For these calculations only $B_1$ is non-zero. (b) Comparison between the analytic predicted and the numerically calculated stresslet $|S|$ over a range of aspect ratios $r_e$. For these calculations only $B_2$ is non-zero.}
\label{Fig:squrimer_comp}
\end{figure}

{\parindent0pt For axisymmetric particles,} we follow the spheroidal squirmers model presented in Ref. \citenum{poehnl20} with slip velocities defined by
\begin{equation}
 v_s(\xi) = \tau_0(\tau_0^2-\xi^2)^{-1/2} P_n^1(\xi).
\end{equation}
$\xi$ and $\tau$ are spheroidal coordinates.
The free space velocity $U_s$ and stresslet $S_0=\frac{3}{2}\sigma_0$ for each individual particle only depends on its aspect ratio $r_e$ and squirming amplitudes $B_n$. In Ref. \citenum{poehnl20}, the figures 3, 8 and 9, \textcolor{black}{are affected by typos in the data generating scripts (more details will be available in a forthcoming erratum)}. The corrected versions are presented here in Fig. \ref{Fig:corr_velo+stress} \textcolor{black}{(and in the erratum)}. The qualitative analysis is unaffected by the miscalculations, however it significantly changes the value of the properties.
In Fig. \ref{Fig:squrimer_comp} we compare the analytic results for the velocity $|U_n|$ and $|S_n|$ induced by $B_1$ and $B_2$ squirmer modes to numerical calculations done with BEM. \textcolor{black}{Eq. 3 in the main text is used to calculate the stresslet.}

\section{Separation distance}

\begin{figure}
\centering 
(a) {\includegraphics[width=0.35\textwidth]{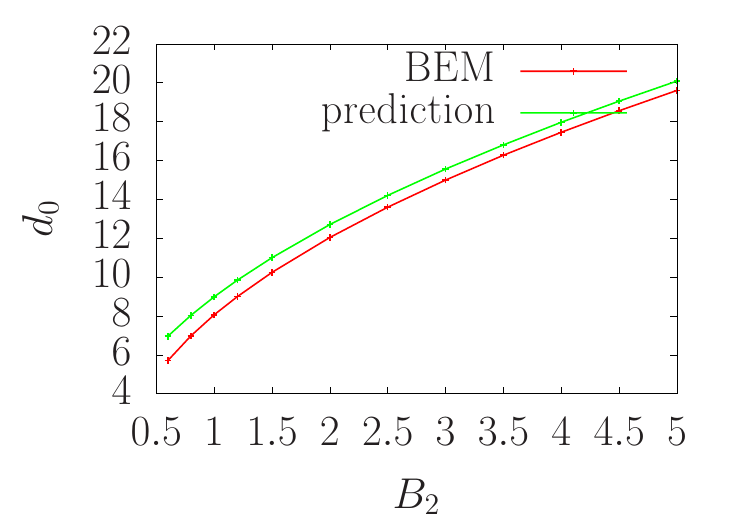}}
 (b) {\includegraphics[width=0.35\textwidth]{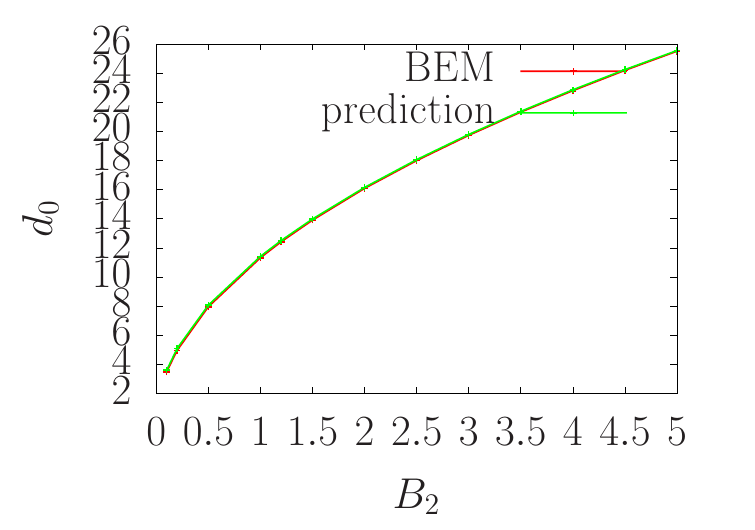}}
 \newline 
 (c) {\includegraphics[width=0.35\textwidth]{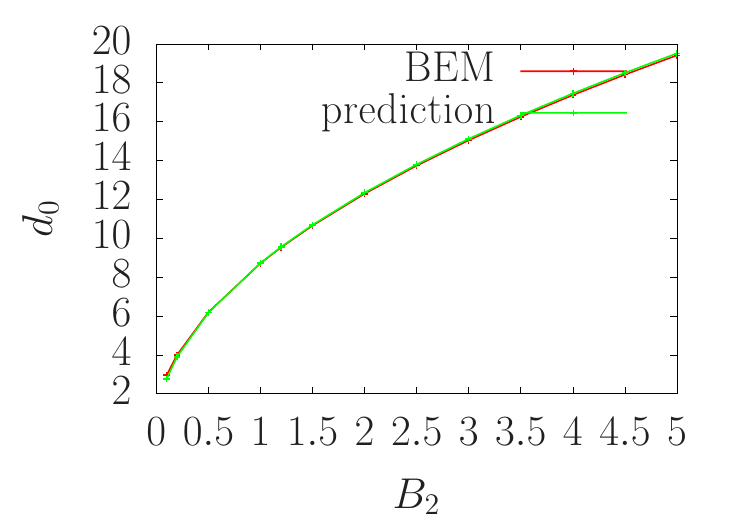}}
\caption{Comparison between the predicted and the calculated (BEM) separation distance for head-to-tail pairs. The velocities $U_s^{(2)}=0.2$ and $U_s^{(1)}=0.25$ are fixed in all cases. We show three different geometries: (a) $r_e^{(1)}=1.1055$ and $r_e^{(2)}=3$, (b) $r_e^{(1)}=1.1055$ and $r_e^{(2)}=1.1055$, and (c) $r_e^{(1)}=0.5774$ and $r_e^{(2)}=0.5774$. %\textcolor{red}{Replace $d$ with $d_0$ in axis labels}
}
\label{Fig:distance}
\end{figure}

{\parindent0pt The second main prediction of our model} besides the stability of the pairs is their \textcolor{black}{steady} separation distance $\textcolor{black}{d_0}$. $\textcolor{black}{d_0}$ can even be calculated for two particles which are only stable against perturbations in the direction of propulsion. In Fig. \ref{Fig:distance} we show how our prediction compares against the distance calculated with the BEM. Similar to the stability analysis in the main text, we generally obtain good agreement between the theory and BEM, \textcolor{black}{especially for the spheres and the prolate particles. For the oblate particles, our analytical prediction are consistently smaller then the results of our numerical calculations, and the relative error becomes more pronounced for small  separation distance $d_0$ between the particles.}

\section{Induced charge electrophoresis}

{\parindent0pt In order to model the interaction of particles with non-axisymmetric actuation,} we first consider a single spheroidal Janus particle moving by ICEP in unbounded solution. The particle consists of a dielectric core with a fraction of its surface covered by metal. The metal coverage is axisymmetric, and the particle aspect ratio is defined as above. A uniform electric field $\mathbf{E} = E_0 \hat{z}$ is perpendicular to the particle axis $\hat{d}$: $\hat{z} = \hat{e}$. In the DC limit, the time-averaged electrostatic potential $\textcolor{black}{\varphi}$ obeys $\nabla^2 \textcolor{black}{\varphi} = 0$, as well as the boundary conditions $\hat{n} \cdot \nabla \textcolor{black}{\varphi} = 0$ on the surface of the particle, and $\textcolor{black}{\varphi} \rightarrow -E_0 z$ as $|\mathbf{x}| \rightarrow \infty$.  As described in Refs. \citenum{kilic11} and \citenum{katuri22}, the surface slip can be obtained from the surface potential. Turning to the dynamics of interacting  pairs, we fix $\mathbf{v}_s^{(\alpha)}$ to be the distribution for an isolated particle, i.e., we neglect electrostatic interactions between particles. Due to the constant surface slip, the particles in this model can be regarded as non-axisymmetric squirmers. The flow  obeys the same equations as for the axisymmetric squirmers.

The coverage of a particle by metal is defined by a parameter $\chi_0 \in [1,1]$. For a spheroid with a its center at the origin and its axis of propulsion $\hat{d}$ aligned with the z-axis, regions on the surface of particle with $z \in [-1,\chi_0]$ are taken to be metal, and regions with $z \in (\chi_0, 1]$ are insulating.  (Here, as discussed in the main text, the semi-axis of propulsion provides the characteristic length, such that the two poles of the particle are at $z = \pm 1$.) Thus, a particle with $\chi_0 = 0$ has half coverage.

In Fig. 2 in the main text, we plot numerical results for oblate ICEP squirmers using circles. These results were obtained for particles with metal coverage slightly more than half: $\chi_0 = 0.5$. This coverage $\chi_0$ was chosen to obtain a finite gap between particles in a bound state for several values of $\Gamma$. For other values of $\chi_0$, particles tend to ``crash'' into each other. (We that in Ref. \citenum{katuri22}, the presence of a nearby bounding surface, i.e., a hard planar wall, was essential for obtaining bound states of half-covered particles.)

\section{Axisymmetric squirmer model}
\begin{figure}
 \includegraphics[width=0.9\textwidth]{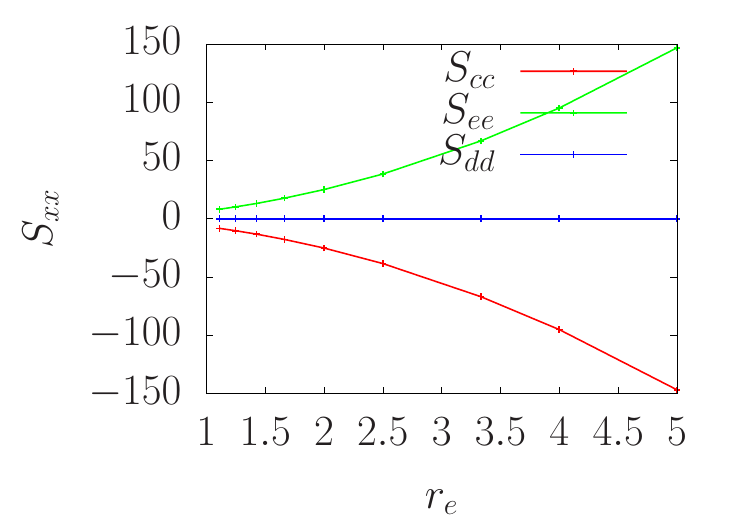}
 \caption{The $S_{dd}$, $S_{cc}$ and $S_{ee}$ contributions for a squirmer with constant asymmetric mode $\tilde{B}=1$ for changing aspect ratio $r_e$.  The lines represent only a guide to the eye.}
\label{Fig:asym_stress}
\end{figure}

{\parindent0pt In Fig. \ref{Fig:asym_stress} we show the three diagonal components of the stresslet} tensor for varying aspect ratio $r_e$. $S_{dd}$ and $U_s$ are zero for all aspect ratios, and $S_{cc}$ and $S_{ee}$ decrease and increase asymmetrically.

\bibliography{spheroidal_squirmers_SI}